\documentclass[twocolumn, showpacs, aps, prb, superscriptaddress,
floatfix]{revtex4}
\usepackage{graphicx}
\usepackage{amsmath}
\begin{document}

\title{A real-space grid implementation of the Projector Augmented Wave method}
\author{J. J. Mortensen}
\author{L. B. Hansen}
\author{K. W. Jacobsen}
\address{CAMP and Department of Physics, Technical University of
  Denmark, DK-2800 Lyngby, Denmark}
\date{\today}

\begin{abstract}
  A grid-based real-space implementation of the Projector Augmented
  Wave (PAW) method of P. E. Bl{\"o}chl [Phys. Rev. B {\bf 50}, 17953
  (1994)] for Density Functional Theory (DFT) calculations is
  presented.  The use of uniform 3D real-space grids for representing
  wave functions, densities and potentials allows for flexible
  boundary conditions, efficient multigrid algorithms for solving
  Poisson and Kohn-Sham equations, and efficient parallelization using
  simple real-space domain-decomposition.  We use the PAW method to
  perform all-electron calculations in the frozen core approximation,
  with smooth valence wave functions that can be represented on
  relatively coarse grids.  We demonstrate the accuracy of the method
  by calculating the atomization energies of twenty small molecules,
  and the bulk modulus
  and lattice constants of bulk aluminum.  We show that the approach
  in terms of computational efficiency is comparable to standard
  plane-wave methods, but the memory requirements are higher.
\end{abstract}
\maketitle

\section{Introduction}

Density Functional Theory\cite{Hoh64, Koh65} (DFT) combined with the
Generalized Gradient Approximation (GGA) for the exchange and
correlation functional has become a popular method for studying
materials and molecules at the atomic scale.  Recently, there has been
an increasing interest in using uniform real-space grids and
finite-difference methods for doing DFT calculations\cite{Che94,
  Sei95, Hos95, Bri96, Ber97, Anc99, Ber00, Bec00, Wan00, Hei01,
  Wag01, Tak01, Mar03, Sch04}.  Real-space grids give an unbiased
description of the wave functions and the quality of the description
can easily be controlled by changing the grid-point density.
Finite-difference operators are used because the wave function values
are given on grid points in real-space, and not in terms of a basis
set.  By doing all operations in real-space, parallelization can be
done by simple domain decomposition\cite{Shi01, Bri96, Liu03}.  Furthermore,
real-space methods can make use of multigrid acceleration
schemes\cite{Bra77} for solving the Kohn-Sham equations\cite{Koh65}
and the Poisson equation.  A further advantage of real space methods
is the possibility for imposing localization constraints on the wave
functions, which is the basis for linearly scaling electronic
structure methods\cite{Shi01, Ber00} (order-$N$ methods).

Today, one of the most used methods for performing DFT calculations is
the pseudo potential method using periodic super-cells and plane-wave
expansions for the pseudo wave-functions.  This method shares with the
grid-based methods the properties of unbiased representation of the
wave functions and simple control of the quality of a calculation (by
changing the number of plane waves).  However, there are three major
difficulties with a plane-wave representation for the wave functions.
1) Working with spatially localized wave functions, which is important
for order-$N$ methods, is difficult with the extended nature of plane
waves.  2) Not all operations involving the wave functions, densities
and potentials can be done directly in the plane-wave representation,
and Fourier transformations to and from real-space must be carried
out.  Transformations between real and reciprocal spaces are highly
non-local operations, and therefore difficult to parallelize.  3) Due
to the periodicity of plane waves, the natural boundary conditions for
a plane wave calculation is periodic boundary conditions.  Although
all three problems have been addressed\cite{Mos02, Gal92, Goe03,
  Blo95, Mar99}, the suggested solutions are not as simple as for
grid-based approaches where all three problems have simple solutions.

An advantage of a plane-wave representation for the wave functions is
its compactness. The memory footprint of a wave function is typically
10 times larger in a real-space grid representation compared to a
plane-wave representation of similar accuracy.  For this reason it is
important to use soft pseudo wave functions that can be accurately
represented on coarse grids.  To our knowledge, until now, all
applications of grid-based electronic structure calculations have used
norm-conserving pseudopotentials.  One way to get smoother pseudo wave
functions is to relax the norm-conservation of the wave functions and
use ultra-soft pseudopotentials\cite{Van90, Laa92} or the Projector
Augmented Wave (PAW) method\cite{Blo94, Blo03}.  We have decided to
use the PAW method.  The PAW method works with soft valence wave
functions and, similar to the ultra-soft pseudopotential method, the
wave functions need not be normalized.  Contrary to the ultra-soft
pseudopotential method, the PAW method is an all-electron method
within the frozen core approximation, giving access to the true wave
functions and the full electron density.  The PAW method has been
implemented for plane waves by several groups\cite{Blo94, Blo03,
  Hol97, Kre99, Val99, Tac01}.

We see the combination of real-space grid-based methods and the PAW
method as an important step towards enabling larger calculations at a
level of accuracy which is essentially all-electron in nature.  There
is a clear trend in electronic structure theory towards larger and
more complex systems as for example nanostructures, large
(bio-)molecular complexes and extended defects in real materials.
Systems which all quickly challenge present day high accuracy DFT
codes, that are typically limited to at most a few hundred atoms.  The
great potential of the method presented here lies in the
parallelization of the real-space algorithms.  This makes it possible
to make use of massively parallel computers as has been demonstrated
by several other groups\cite{Shi01, Bri96, Liu03}.  In this paper, we
focus on how to do accurate DFT calculations efficiently by using a
real-space PAW method.  We demonstrate the accuracy of our grid-based
PAW calculations by showing that we are able to reproduce results for
atomization energies from all-electron DFT calculations.  This very
stringent test shows that the methodology that we have developed is
useful for real applications.

The solution of Poisson's equation is straightforward using multigrid
methods\cite{Bra77} (no Fourier transformations required).  Solving
the Kohn-Sham equations using multigrid methods is a much more
difficult task: Keeping the different eigenstates separated and
orthogonal to each other can be a problem, and representing the
Hamiltonian on the coarse grids can also be problematic.  We have
decided to use the techniques typically used in ``state of the art''
pseudopotential plane-wave calculations\cite{Kre96} as they have been
developed and improved over the the last few decades.  For iteratively
solving the Kohn-Sham equations, we use Pulay mixing techniques for
obtaining the self-consistent density\cite{Pul80, Kre96}, subspace
diagonalizations, and the Residual Minimization Method\cite{Woo85,
  Kre96} using preconditioning of the electronic gradients for
iteratively improving the wave functions.  The preconditioning
operation is a single multigrid V-cycle using only the kinetic energy
operator as an approximate Hamiltonian\cite{Bri96}.

In the following section, we will briefly summarize the PAW method.
Then, in section~\ref{grids}, we will go through the details of our
grid-based formulation of the PAW formalism.  In
section~\ref{groundstate}, we describe how we solve the Kohn-Sham
equations, and the evaluation of atomic forces is discussed in
section~\ref{forces}.  Section~\ref{generalizations} describes
generalizations of the method to periodic systems with use of Brillouin
zone sampling.  In section~\ref{applications} we apply the methodology
that we have developed to a number of example systems and discuss
approximations necessary for realistic calculations.  Finally, in
section~\ref{performance} we discuss the computational performance of
our implementation.

\section{The Projector Augmented Wave method}
\label{PAW}
The notation we use is close to the one used by P. E. Bl{\"o}chl in his
papers on the PAW method\cite{Blo94, Blo03}.  We have used Hartree
atomic units ($\hbar=m=e=1$) and we write the equations for the case
of a spin-paired and finite system of electrons.

The PAW method is based on a transformation between smooth pseudo wave
functions, $\tilde{\psi}_n$, and the true all-electron Kohn-Sham wave
functions, $\psi_n$ ($n$ is the band index).  The core states of the
atoms, $\phi_i^{a,\text{core}}$, are frozen.  Here $a$ is an atom
index and $i$ is a combination of principal, angular momentum, and
magnetic quantum numbers respectively ($n$, $\ell$ and $m$).

Given a smooth pseudo wave function, the corresponding all-electron
wave function, which is orthogonal to the set of
$\phi_i^{a,\text{core}}$ orbitals, can be obtained through a linear
transformation

\begin{equation}
  \psi_n(\mathbf{r}) = \hat{\mathcal{T}}\tilde{\psi}_n(\mathbf{r}).
\end{equation}
The transformation operator, $\hat{\mathcal{T}}$, is given in terms of atom centered
all-electron wave functions, $\phi_i^a(\mathbf{r})$, the
corresponding smooth partial waves, $\tilde{\phi}_i^a(\mathbf{r})$, 
and projector functions, $\tilde{p}_i^a(\mathbf{r})$, as

\begin{equation}
  \hat{\mathcal{T}} = 1 +  \sum_a \sum_i 
  (|\phi_i^a\rangle - |\tilde{\phi}_i^a\rangle)\langle\tilde{p}_i^a|.
  \label{transform}
\end{equation}
The atom centered all-electron wave functions are taken from a
calculation of a single atom with spherical symmetry:
$\phi_i^a(\mathbf{r}) = \phi_{n\ell}^a(r) Y_L(\hat{\mathbf{r}})$,
where the $Y_L$'s are real-valued spherical harmonics ($L$ is a
combined index for $\ell$ and $m$).

A radial cutoff distance, $r_c^a$, defining the atomic augmentation
sphere is chosen.  This radius is similar to a cutoff radius for a
pseudopotential.  The larger the augmentation sphere, the smoother
the pseudo wave functions, but overlap with neighboring augmentation
spheres must be avoided.

For all all-electron valence states smooth partial waves
$\tilde{\phi}_i^a(\mathbf{r}) = \tilde{\phi}_{n\ell}^a(r)
Y_L(\hat{\mathbf{r}})$ are constructed.  The partial waves must match
the corresponding all-electron waves for $r > r_c^a$.  In this way,
the correction in parenthesis in Eq.~(\ref{transform}) is zero outside
the augmentation spheres and we will have $\hat{\mathcal{T}} = 1$ in
this region.  Notice, that there are no norm-conservation requirements
to meet when choosing the shape of $\tilde{\phi}_{n\ell}^a(r)$ inside
the augmentation sphere.

Smooth projector functions must also be defined --- one for each
partial wave: $\tilde{p}_i^a(\mathbf{r}) = \tilde{p}_{n\ell}^a(r)
Y_L(\hat{\mathbf{r}})$.  They must be localized inside the
augmentation spheres and satisfy $\langle \tilde{p}_{i_1}^a |
\tilde{\phi}_{i_2}^a \rangle = \delta_{i_1i_2}$, which for the radial
part gives
\begin{equation}
  \int_o^{r_c^a} r^2 dr
  \tilde{p}_{n\ell}^a(r) \tilde{\phi}_{n'\ell}^a(r)
  = \delta_{nn'}.
\end{equation}
With this construction we have $\hat{\mathcal{T}}
\tilde{\phi}_i^a(\mathbf{r}) = \phi_i^a(\mathbf{r})$.

In principle, an infinite number of projectors and partial waves are
required for the PAW method to be exact.  For practical calculations,
a high accuracy data set will need only one or two projector functions
for each angular momentum channel of importance.  This is similar to
an ultra-soft pseudopotential, where a comparable number of projectors is
needed.  One partial wave is usually taken as the bound valence state,
and additional waves can be taken from ``excited states'' ---
solutions to the radial Kohn-Sham equation at different non-eigenvalue
energies.  The construction of partial waves and projector functions
is described in appendix~\ref{construction}.

\subsection{PAW densities}

>From the atomic frozen core electron density, $n_c^a(r)$, a smooth
core electron density, $\tilde{n}_c^a(r)$, is constructed.  It must be
identical to $n_c^a(r)$ for $r > r_c^a$.  There is no norm-conservation
requirement to meet when choosing the shape of $\tilde{n}_c^a(r)$
inside the augmentation sphere.

The pseudo electron density has contributions from the wave functions
and from the atom centered smooth core electron densities:
\begin{equation}
  \tilde{n}(\mathbf{r}) = \sum_n f_n |\tilde{\psi}_n(\mathbf{r})|^2 +
  \sum_a \tilde{n}_c^a(|\mathbf{r} - \mathbf{R}^a|),
  \label{nofr}
\end{equation}
where the $f_n$'s are occupation numbers and $\mathbf{R}^a$ is the
position of atom $a$.

An atomic density matrix (see Eq.~(22) in Ref.~\onlinecite{Blo03}) is
defined as
\begin{equation}
  D_{i_1i_2}^a = \sum_n \langle \tilde{p}_{i_1}^a | \tilde{\psi}_n
  \rangle f_n 
\langle \tilde{\psi}_n | \tilde{p}_{i_2}^a \rangle
\end{equation}
where
\begin{equation}
  P_{ni}^a = \langle \tilde{p}_i^a | \tilde{\psi}_n \rangle =
  \int d\mathbf{r} 
  \tilde{p}_i^a(\mathbf{r} - \mathbf{R}^a)
  \tilde{\psi}_n(\mathbf{r})
\end{equation}

The PAW formalism defines atom centered all-electron and pseudo
electron densities as
\begin{equation}
  n^a(\mathbf{r}) = \sum_{i_1i_2} D_{i_1i_2}^a
  \phi_{i_1}^a(\mathbf{r}) \phi_{i_2}^a(\mathbf{r}) + 
  n_{\text{c}}^a(r)
  \label{n}
\end{equation}
and
\begin{equation}
  \tilde{n}^a(\mathbf{r}) = \sum_{i_1i_2} D_{i_1i_2}^a
  \tilde{\phi}_{i_1}^a(\mathbf{r}) \tilde{\phi}_{i_2}^a(\mathbf{r}) +
  \tilde{n}_{\text{c}}^a(r),
  \label{nt}
\end{equation}
respectively.  By construction, $n^a(\mathbf{r} - \mathbf{R}^a)$ is
identical to the all-electron density, $n(\mathbf{r})$, for
$|\mathbf{r} - \mathbf{R}^a| < r_c^a$ and $\tilde{n}^a(\mathbf{r} -
\mathbf{R}^a) = \tilde{n}(\mathbf{r})$ for $|\mathbf{r} -
\mathbf{R}^a| < r_c^a$ (see Ref.~\onlinecite{Blo94} for details).
Therefore, the true all-electron density can be obtained from the
pseudo electron density:
\begin{equation}
  n(\mathbf{r}) = \tilde{n}(\mathbf{r}) 
  + \sum_a [n^a(\mathbf{r} -
  \mathbf{R}^a) - \tilde{n}^a(\mathbf{r} - \mathbf{R}^a)].
\end{equation}
Again, the correction is zero outside the augmentation spheres.

A neutral charge density, $\tilde{\rho}(\mathbf{r})$, is obtained by
adding compensation charges, $\tilde{Z}^a(\mathbf{r})$, inside the
augmentation spheres to the pseudo electron density.  These charges
compensate for the lack of norm-conservation and for the nuclear
charge:
\begin{equation}
  \tilde{\rho}(\mathbf{r}) = \tilde{n}(\mathbf{r})
  + \sum_a \tilde{Z}^a(\mathbf{r} - \mathbf{R}^a).
  \label{rhot}
\end{equation}

Using localized functions $\tilde{g}_L^a(\mathbf{r}) =
\tilde{g}_\ell^a(r) Y_L(\hat{\mathbf{r}})$ normalized as
\begin{equation}
  \int d \mathbf{r} r^\ell \tilde{g}^a_L(\mathbf{r})
  Y_L(\hat{\mathbf{r}}) = 1,
  \label{norm}
\end{equation}
the compensation charges are constructed with electrostatic multipole
moments $Q_L^a$:
\begin{equation}
  \tilde{Z}^a(\mathbf{r}) = \sum_L Q_L^a \tilde{g}_L^a(\mathbf{r}).
  \label{Zt}
\end{equation}

The values to be used for the electrostatic multipole moments, $Q_L^a$,
are found by requiring the pseudo charge density, $\tilde{n}^a +
\tilde{Z}^a$, to have the same electrostatic multipole moments as the
all-electron charge density, $n^a + Z^a$, where $Z^a(\mathbf{r}) =
-\mathcal{Z}^a \delta(\mathbf{r})$ is the nuclear charge density
($\mathcal{Z}^a$ is the atomic number).  This requirement can be
expressed as
\begin{equation}
  \int d \mathbf{r} r^\ell 
  [\tilde{n}^a(\mathbf{r}) + \tilde{Z}^a(\mathbf{r}) - 
  n^a(\mathbf{r}) - Z^a(\mathbf{r})]
  Y_L(\hat{\mathbf{r}}) = 0.
\end{equation}

Inserting Eqs.~(\ref{n}), (\ref{nt}) and (\ref{Zt}) we get:
\begin{equation}
  Q_L^a = \sum_{i_1i_2} \Delta_{Li_1i_2}^a D_{i_1i_2}^a + 
  \Delta^a \delta_{\ell 0},
  \label{Q}
\end{equation}
where the constants $\Delta_{Li_1i_2}^a$ and $\Delta^a$ are given by:
\begin{equation}
  \Delta_{Li_1i_2}^a = \int d\mathbf{r} Y_L(\hat{\mathbf{r}}) r^\ell
  [\phi_{i_1}^a(\mathbf{r})\phi_{i_2}^a(\mathbf{r}) -
   \tilde{\phi}_{i_1}^a(\mathbf{r})\tilde{\phi}_{i_2}^a(\mathbf{r})]
\end{equation}
and 
\begin{equation}
  \Delta^a = \int d\mathbf{r} Y_{00}(\hat{\mathbf{r}})
  [-\mathcal{Z}^a\delta(\mathbf{r}) + n_c^a(r) - \tilde{n}_c^a(r)].
\end{equation}

\subsection{The PAW total energy}

The PAW total energy is a function of the pseudo wave functions,
$\tilde{\psi}_n(\mathbf{r})$, and the occupation numbers, $f_n$.  The
energy can be divided into a ``soft'' contribution, $\tilde{E}$, and
corrections for each atom (see Ref.~\onlinecite{Blo94} and
\onlinecite{Blo03}):
\begin{equation}
  E = \tilde{E} + \sum_a (E^a - \tilde{E}^a).
\end{equation}

The ``soft'' energy contribution is:
\begin{eqnarray}
  \tilde{E} & = & 
  \sum_n f_n \int d\mathbf{r} 
  \tilde{\psi}_n^*(\mathbf{r}) (-\tfrac{1}{2} \nabla^2)
  \tilde{\psi}_n(\mathbf{r})\nonumber \\
  & & \mbox{} + \tfrac{1}{2} \int d\mathbf{r} 
  \tilde{v}^{\text{H}}(\mathbf{r}) \tilde{\rho}(\mathbf{r})
  + E_{\text{xc}}[\tilde{n}(\mathbf{r})]\nonumber \\
  & & + \int d\mathbf{r} \tilde{n}(\mathbf{r})
  \sum_a \bar{v}^a(|\mathbf{r} - \mathbf{R}^a|),
\end{eqnarray}
where $\tilde{v}^{\text{H}}(\mathbf{r})$ is the pseudo Hartree potential,
satisfying Poisson's equation $\nabla^2 \tilde{v}^{\text{H}} = -4 \pi
\tilde{\rho}$ and $E_{\text{xc}}$ is an exchange-correlation
functional.  Finally, $\bar{v}^a(r)$ is an arbitrary
localized potential vanishing for $r > r_c^a$.  The soft energy
contribution, $\tilde{E}$, is to be evaluated on three dimensional
grids in real space.

The atomic corrections to the energy ($E^a - \tilde{E}^a$) are given
by:
\begin{align}
  E^a & =
  \sum_i^{\text{core}}
  \int d \mathbf{r} \phi_i^{a,\text{core}}(\mathbf{r}) (-\tfrac{1}{2} \nabla^2)
  \phi_i^{a,\text{core}}(\mathbf{r}) \nonumber \\
  & + \sum_{i_1i_2} D_{i_1i_2}^a
  \int d \mathbf{r} \phi_{i_1}^a(\mathbf{r}) (-\tfrac{1}{2} \nabla^2)
  \phi_{i_2}^a(\mathbf{r})\nonumber \\
  & + \tfrac{1}{2} \int d \mathbf{r} \int d \mathbf{r}'
  \frac{[n^a(\mathbf{r}) + Z^a(\mathbf{r})]
    [n^a(\mathbf{r}') + Z^a(\mathbf{r}')]}
  {|\mathbf{r} - \mathbf{r}'|}\nonumber \\
  & + E_{\text{xc}}[n^a(\mathbf{r})]
  \label{ea}
\end{align}
and
\begin{align}
  \tilde{E}^a & = 
  \sum_{i_1i_2} D_{i_1i_2}^a
  \int d \mathbf{r} \tilde{\phi}_{i_1}^a(\mathbf{r}) (-\tfrac{1}{2} \nabla^2 )
  \tilde{\phi}_{i_2}^a(\mathbf{r})\nonumber \\
  & + \tfrac{1}{2} \int d \mathbf{r} \int d \mathbf{r}'
  \frac{[\tilde{n}^a(\mathbf{r}) + \tilde{Z}^a(\mathbf{r})]
    [\tilde{n}^a(\mathbf{r}') + \tilde{Z}^a(\mathbf{r}')]}
  {|\mathbf{r} - \mathbf{r}'|}\nonumber \\
  & + E_{\text{xc}}[\tilde{n}^a(\mathbf{r})]
  + \int d \mathbf{r} \tilde{n}^a(\mathbf{r}) \bar{v}^a(r).
  \label{eat}
\end{align}
The energy contributions for $E^a$ and $\tilde{E}^a$, are evaluated on
radial grids inside the augmentation spheres.

By using Eqs.~(\ref{n}), (\ref{nt}), (\ref{Zt}) and (\ref{Q}) we can
reduce the atomic correction, $E^a - \tilde{E}^a$, to a function of
$D_{i_1i_2}^a$:
\begin{align}
  E^a - \tilde{E}^a = &
  A^a + \sum_{i_1i_2} B_{i_1i_2}^a D_{i_1i_2}^a \nonumber \\
  & + \sum_{i_1i_2} \sum_{i_3i_4}
  D_{i_1i_2}^a C_{i_1i_2i_3i_4}^a D_{i_3i_4}^a \nonumber \\
  & + \Delta E_{xc}^a(\{D_{i_1i_2}^a\}).
  \label{ABC}
\end{align}
The constants $A^a$, $B_{i_1i_2}^a$ and $C_{i_1i_2i_3i_4}^a$ are
evaluated in the appendix~\ref{constants}.  

The last term is an exchange-correlation correction:
\begin{equation}
  \Delta E_{xc}^a(\{D_{i_1i_2}^a\}) = E_{\text{xc}}[n^a(\mathbf{r})]
    - E_{\text{xc}}[\tilde{n}^a(\mathbf{r})],
    \label{xca}
\end{equation}
which is a function of $D_{i_1i_2}^a$ through Eqs.~(\ref{n}) and (\ref{nt}).
For local and semi-local exchange-correlation functionals, $\Delta
E_{xc}^a(\{D_{i_1i_2}^a\})$ is written as an
integration inside the augmentation sphere.
There are several possibilities for the evaluation of this term.  We
use radial integration along lines from the center to a number of
points distributed evenly on the surface of the augmentation
sphere\cite{Hol97, Fli99}.  We find that this approach is the simplest for
GGA functionals\cite{simple}.  Alternatively, one can expand
the atomic densities in spherical harmonics, as described in
Ref.~\onlinecite{Blo94} or use grid free approaches\cite{Zhe96, Ber98}.

\section{Uniform 3D grids}

\label{grids}
In this section we give the details of our real-space grid-based
implementation of the PAW method.  From now on, wave functions,
electron densities and potentials are represented on three-dimensional
uniform grids in real-space.

\subsection{Localized functions and the double grid technique}

In a grid representation, integrals over space are turned into sums
over grid points.  In the PAW method we often need to calculate the
integral of a localized function, centered on an atom, multiplied by a
function extended over all of space.  As an example, let us take a
projector function, $\tilde{p}_i^a(\mathbf{r} - \mathbf{R}^a)$, centered
on atom $a$ at position $\mathbf{R}^a$, multiplied by an extended
wave function, $\tilde{\psi}_n(\mathbf{r})$.

In the following, we use the index $G$ to index the grid points used for the
wave functions.  Transforming the integral to a sum over grid points,
$G$, with $V_c$ being the volume per grid point, we get:
\begin{equation}
  P_{ni}^a = V_c \sum_G \tilde{p}_{iG}^a, \tilde{\psi}_{nG},
  \label{sum}
\end{equation}
where $\tilde{\psi}_{nG} = \tilde{\psi}_n(\mathbf{r}_G)$ and
$\mathbf{r}_G$ is the position of grid point $G$ (only the grid points
in the localized region around atom $a$ needs to be summed over).
For $\tilde{p}_{iG}^a$ we could use
$\tilde{p}_i^a(\mathbf{r}_G - \mathbf{R}^a)$.  However, this is not
accurate enough, unless we use a very fine grid, which would
compromise efficiency.  Instead we use the elegant double grid
technique of Ono and Hirose\cite{Ono99}.  Here, the extended function
is interpolated to a finer grid with grid
points $f$: $\tilde{\psi}_{nf} = \sum_G I_{fG} \tilde{\psi}_{nG}$.
The interpolation operator $I_{fG}$ takes the wave function from a
coarse grid to a fine grid.  Typically, a cubic interpolation is used
and the fine grid has five times more points in each direction as the
coarse grid\cite{Ono99, Tak01}.  The localized projector function is
evaluated on a fine grid as $\tilde{p}_{if}^a =
\tilde{p}_i^a(\mathbf{r}_f - \mathbf{R}^a)$.  With $v$ being the
volume per fine grid point we get a more accurate sum:
\begin{eqnarray}
  v \sum_f \tilde{p}_{if}^a \tilde{\psi}_{nf} = 
  v \sum_f \tilde{p}_{if}^a \sum_G I_{fG} \tilde{\psi}_{nG} \nonumber \\
  = V_c \sum_G [(v / V_c) \sum_f I_{fG} \tilde{p}_{if}^a] \tilde{\psi}_{nG}.
\end{eqnarray} 

Comparing the rightmost expression with Eq.~(\ref{sum}), we identify
the term in square brackets as the more accurate expression for
$\tilde{p}_{iG}^a$:
\begin{equation}
  \tilde{p}_{iG}^a = \frac{v}{V_c} \sum_f I_{fG}
  \tilde{p}_i^a(\mathbf{r}_f - \mathbf{R}^a).
\end{equation} 

This is equivalent to a restriction operation taking the localized
function from the fine temporary grid to the coarse grid
(restriction is the opposite of interpolation).  Notice that in
Eq.~(\ref{sum}), the sum is over coarse grid points, and no actual
interpolation of the extended function needs to be done.  The
evaluation of the localized function on the temporary grid is a
relatively inexpensive operation, with an operation count which for
each atom is independent of the system size.

We use the double grid technique to transfer localized functions, such
as $\tilde{p}_i^a(\mathbf{r})$, $\tilde{n}_c^a(r)$ and $\bar{v}^a(r)$,
to values on real-space grids ($\tilde{p}_{iG}^a$, $\tilde{n}_{cG}^a$
and $\bar{v}_g^a$).

\subsection{Real-space grid formulation of the PAW method}

The formulas for densities, potentials, and energies, given in
section~\ref{PAW}, must be translated to a discretized form for use
with a discrete representation of wave functions, densities and
potentials.  The pseudo electron density, Eq.~(\ref{nofr}), is calculated on a coarse grid as
\begin{equation}
  \tilde{n}_G = \sum_n f_n |\tilde{\psi}_{nG}|^2 +
  \sum_a \tilde{n}_{cG}^a.
\end{equation}

The smooth atomic core electron densities can be chosen very soft, so
that they can be added to the coarse grid.  

The pseudo electron
density on the coarse grid is interpolated to a finer grid (grid
points indexed by $g$):
\begin{equation}
  \tilde{n}_g = \sum_G I_{gG} \tilde{n}_G.
\end{equation}
We use a cubic interpolation for $I_{gG}$, and the fine grid has twice
as many grid points as the coarse grid in each direction.  From the
pseudo electron density on the fine grid, one can obtain the neutral
charge density as [see Eq.~(\ref{rhot})]:
\begin{equation}
  \tilde{\rho}_g = \tilde{n}_g + \sum_a \tilde{Z}_g^a.
\label{rho}
\end{equation}

Using our grid representation for wave functions, densities and potentials, we get for
$\tilde{E}$:
\begin{eqnarray}
  \tilde{E} & = & 
  \sum_n f_n V_c \sum_G
  \tilde{\psi}_{nG}^* \sum_{G'} (-\tfrac{1}{2} L_{GG'})
  \tilde{\psi}_{nG'}\nonumber \\
  & & \mbox{} + \tfrac{1}{2} V_f \sum_g
  \tilde{v}_g^{\text{H}} \tilde{\rho}_g
  + E_{\text{xc}}(\{\tilde{n}_g\}, V_f)\nonumber \\
  & & \mbox{} + V_f \sum_g \tilde{n}_g \sum_a \bar{v}_g^a,
\end{eqnarray}
where $V_c$ and $V_f$ are the volumes per grid point for the
coarse wave function grids and the fine density and potential grids
respectively, and $L_{GG'}$ is a central finite
difference representation of the Laplacian.  The discretization that
we use for the Laplacian, uses a total of twelve neighbor points,
giving an error of the order of $h^6$, where $h = V_c^{1/3}$ is the
grid spacing.  

The pseudo Hartree potential $\tilde{v}_g^{\text{H}}$ is found by
solving Poisson's equation, $\nabla^2 \tilde{v}^{\text{H}} = -4 \pi
\tilde{\rho}$, using a discretization for the
Laplacian:
\begin{equation}
  \sum_{g'} C_{gg'} \tilde{v}_{g'}^{\text{H}} =
  -4 \pi \tilde{\rho}_g.
\end{equation}

This equation is solved using the multigrid technique pioneered by
Brandt\cite{Bra77}.  Solving the Poisson equation iteratively on the
finest grid will quickly reduce the short wavelength errors, but
errors with larger wavelengths compared to the grid spacing are reduced
only slowly.  The multigrid technique introduces a series of coarser
grids where the long wavelength errors can be effectively reduced.

We use a Mehrstellen discretization\cite{Bri96}, where $C_{gg'} =
(\mathbf{B}^{-1}\mathbf{A})_{gg'}$ is expressed in terms of two
short-ranged finite difference operators $A_{gg'}$ and $B_{gg'}$:
\begin{equation}
  \sum_{g'} A_{gg'} \tilde{v}_{g'}^{\text{H}} =
  -4 \pi \sum_{g'} B_{gg'} \tilde{\rho}_{g'}.
\end{equation}

For the coarse grids used in the multigrid V-cycle, 
simple nearest neighbor central finite-difference Laplacians are used.

\subsection{Soft compensation charges}

Adding the compensation charges, $\tilde{Z}^a$, to the pseudo electron
density [Eqs.~(\ref{rhot}) and (\ref{rho})] will require a very fine
density grid in order to get an accurate description of the charge.
The problem is that the compensation charges must be localized inside
the augmentation spheres: $\tilde{g}_L^a(\mathbf{r}) = 0$ for $r >
r_c^a$.  Bl{\"o}chl has described a method for plane-wave basis
sets\cite{Blo94}, that allows the use of softer compensation charges
extending outside the augmentation spheres.  We use the same method
for our grid-based approach.  A cutoff radius $\hat{r}_c^a$ larger
than $r_c^a$ is chosen.  Soft compensation charges with the same
electrostatic multipole moments as the localized compensation charges
are constructed:
\begin{equation}
  \hat{Z}^a(\mathbf{r}) = \sum_L Q_L^a \hat{g}_L^a(\mathbf{r}),
\end{equation}
where $\hat{g}_L^a(\mathbf{r})$ is a soft function localized within $r
< \hat{r}_c^a$.  The soft function, $\hat{g}_L^a(\mathbf{r})$, is
normalized in the same way as $\tilde{g}_L^a(\mathbf{r})$ --- see
Eq.~(\ref{norm}).

Equation (\ref{rho}) must now be replaced by:
\begin{equation}
  \tilde{\rho}_g = \tilde{n}_g + \sum_a \hat{Z}_g^a
  = \tilde{n}_g + \sum_a \sum_L Q_L^a \hat{g}_{Lg}^a,
\end{equation}
and a correction must be added to $\tilde{E}$. The
correction is (see Ref.~\onlinecite{Blo94}):
\begin{equation}
  \int d \mathbf{r} \tilde{n}(\mathbf{r}) 
  \sum_a \hat{v}^a(\mathbf{r} - \mathbf{R}^a)
  + \sum_{aa'} U^{aa'},
  \label{correction}
\end{equation}
where $\hat{v}^a(\mathbf{r}) = \sum_L Q_L^a \hat{v}_L^a(\mathbf{r})$, and
\begin{equation}
  \hat{v}_L^a(\mathbf{r}) = \int d \mathbf{r}' 
  \frac{\tilde{g}_L^a(\mathbf{r}') - \hat{g}_L^a(\mathbf{r})}
  {|\mathbf{r}-\mathbf{r}'|}.
\end{equation}

The first term in Eq.~(\ref{correction}) is evaluated on the grid as
$V_f \sum_g \tilde{n}_g \sum_a \hat{v}_g^a$.  The transformation of
the localized potential, $\hat{v}^a(\mathbf{r})$, and the localized
compensation charge, $\hat{Z}^a(\mathbf{r})$, (both vanishing for $r >
\hat{r}_c^a$) to values at grid points, $\hat{v}_g^a$ and
$\hat{Z}_g^a$, is done using the double grid technique\cite{Ono99}.
The last term in Eq.~(\ref{correction}) is a pair potential with range
$\hat{r}_c^a + \hat{r}_c^{a'}$:
\begin{align}
  U^{aa'} & = \tfrac{1}{2} \int d \mathbf{r} \int d \mathbf{r}' 
  \left(
  \frac{\tilde{Z}^a(\mathbf{r} - \mathbf{R}^a)
    \tilde{Z}^{a'}(\mathbf{r}' - \mathbf{R}^{a'})}
  {|\mathbf{r}-\mathbf{r}'|}\right.\nonumber \\
  & - 
  \left.\frac{\hat{Z}^a(\mathbf{r} - \mathbf{R}^a)
    \hat{Z}^{a'}(\mathbf{r}' - \mathbf{R}^{a'})}
  {|\mathbf{r}-\mathbf{r}'|}\right)\nonumber \\
  & = \tfrac{1}{2} \sum_{LL'} Q_L^a V_{LL'}^{aa'} Q_{L'}^{a'},
\end{align}
where
\begin{eqnarray}
  V_{LL'}^{aa'} & = & \int d \mathbf{r} \int d \mathbf{r}' 
  \left(
  \frac{\tilde{g}_L^a(\mathbf{r} - \mathbf{R}^a)
    \tilde{g}_{L'}^{a'}(\mathbf{r}' - \mathbf{R}^{a'})}
  {|\mathbf{r}-\mathbf{r}'|}\right.\nonumber \\
  & & - 
  \left.\frac{\hat{g}_L^a(\mathbf{r} - \mathbf{R}^a)
    \hat{g}_{L'}^{a'}(\mathbf{r}' - \mathbf{R}^{a'})}
  {|\mathbf{r}-\mathbf{r}'|}\right).
\label{pair}
\end{eqnarray}
The pair potential terms, $V_{LL'}^{aa'}$, are functions of the
difference vectors $\mathbf{R}^a - \mathbf{R}^{a'}$.

\subsection{Orthogonality}

The orthogonality constraint of the all-electron wave functions,
$\langle\psi_n|\psi_{n'}\rangle = \delta_{nn'}$, can be expressed in
terms of the pseudo wave functions\cite{Blo94} as
$\langle\tilde{\psi}_n|\hat{O}|\tilde{\psi}_{n'}\rangle =
\delta_{nn'}$, where the PAW overlap operator $\hat{O}$ is non-local:
\begin{equation}
  \hat{O} = 1 + \sum_a \sum_{i_1i_2}
  |\tilde{p}_{i_1}^a\rangle O_{i_1i_2}^a \langle\tilde{p}_{i_2}^a|,
\end{equation}
with 
\begin{eqnarray}
  O_{i_1i_2}^a & = & \int d\mathbf{r} 
  [\phi_{i_1}^a(\mathbf{r})\phi_{i_2}^a(\mathbf{r}) -
  \tilde{\phi}_{i_1}^a(\mathbf{r})\tilde{\phi}_{i_2}^a(\mathbf{r})]\nonumber \\
  & = & \sqrt{4 \pi} \Delta_{00i_1i_2}^a.
\end{eqnarray}
The discretised overlap operator looks like:
\begin{equation}
  O_{GG'} = \delta_{GG'} + V_c \sum_a \sum_{i_1i_2}
  \tilde{p}_{i_1G}^a O_{i_1i_2}^a \tilde{p}_{i_2G'}^a,
\end{equation}
and the orthogonality
constraint of the pseudo wave functions can be expressed as
\begin{equation}
  V_c \sum_{GG'} \tilde{\psi}_{nG}^* O_{GG'} 
  \tilde{\psi}_{n'G'} = \delta_{nn'}.
\label{O}
\end{equation}

\section{Finding the ground state}

\label{groundstate}
In order to find the electronic ground state it is necessary to
calculate the derivatives of the total energy with respect to the
wave function values.  This ``electronic gradient'' can be expressed in terms
of a Hamiltonian $H_{GG'}$:
\begin{equation}
  \frac{1}{V_c}
  \frac{\partial E}{\partial\tilde{\psi}_{nG}^*}
  = f_n \sum_{G'} H_{GG'} \tilde{\psi}_{nG'}.
\label{grad}
\end{equation}

The Hamiltonian is given as a sum of the kinetic energy operator and
the local and non-local parts of the effective potential:
\begin{equation}
  H_{GG'} = -\tfrac{1}{2} L_{GG'} + \tilde{v}_G \delta_{GG'} + 
  V_c \sum_a \sum_{i_1i_2} \tilde{p}_{i_1G}^a H_{i_1i_2}^a
  \tilde{p}_{i_2G'}^a.
  \label{H}
\end{equation}

Explicit formulas for the local effective potential,
$\tilde{v}_G$, and the ``atomic'' Hamiltonian, $H^a_{i_1i_2}$, are
given in the appendix~\ref{ham}.

The set of orthonormalized ground-state wave functions that diagonalize
the Hamiltonian matrix $H_{nn'} = V_c \sum_{GG'} \tilde{\psi}_{nG}^*
H_{GG'} \tilde{\psi}_{n'G'}$, must satisfy the Kohn-Sham equations:
\begin{equation}
  \sum_{G'} (H_{GG'} - \epsilon_n O_{GG'}) \tilde{\psi}_{nG'} = 0.
\label{KS}
\end{equation}

\subsection{Residual minimization method and Pulay mixing}

In order to locate the self-consistent ground state, we use the Residual
Minimization Method of Wood and Zunger\cite{Woo85} (see also
Ref.~\onlinecite{Kre96}).  The residuals are calculated as
\begin{equation}
  R_{nG} = \sum_{G'} (H_{GG'} - \epsilon_n O_{GG'}) \tilde{\psi}_{nG'},
\end{equation}
where $\epsilon_n$ is the current estimate of the eigenvalue of the
$n$'th band.  Using a preconditioning operator $\hat{P}$ (to be
discussed later), we can improve the wave function by taking a step
along the direction of the preconditioned residual: $\tilde{\psi}_{nG}
+ \lambda \hat{P} R_{nG}$ ($\lambda$ is the step length).  The
optimal step length is found by minimizing the norm of the residual
for the new guess:
\begin{eqnarray}
  R_{nG}' & = \sum_{G'} (H_{GG'} - \epsilon_n O_{GG'}) 
  (\tilde{\psi}_{nG'} + \lambda \hat{P} R_{nG'})\nonumber \\
  & = R_{nG} + \lambda \sum_{G'} (H_{GG'} - \epsilon_n O_{GG'}) 
  \hat{P} R_{nG'}.
\end{eqnarray}
Finding the optimal value for $\lambda$ amounts to finding the minimum
of a second order polynomial in $\lambda$.
Having found the optimal step length for the first step, we do the
actual update of the wave function by taking an additional step using
the same step length in the direction of the preconditioned residual
$R_{nG}'$:
\begin{equation}
  \tilde{\psi}_{nG} \leftarrow \tilde{\psi}_{nG} 
  + \lambda \hat{P} R_{nG}
  + \lambda \hat{P} R_{nG}'.
\end{equation}
For updating one wave function, one must apply the Hamiltonian twice,
and two preconditioning operations are required.

If we were to take steps along the residual (and not along the
preconditioned residual), we would need very many
iterations in order to converge.  The problem is that the residual
vector is not necessarily parallel to the error vector (which we don't
know).  The purpose of preconditioning is to produce a direction that more
accurately represents the error vector\cite{Pay92}.

We would get the optimal preconditioned residual $\tilde{R}_n =
\hat{P} R_n$ by solving $(\hat{H} - \epsilon_n \hat{O}) \tilde{R}_n =
R_n$.  Instead of solving this equation exactly, we solve
approximately the simpler equation $-\tfrac{1}{2}\nabla^2 \tilde{R}_n
= R_n$.  This is done using one multigrid V-cycle, where a nearest
neighbor discretization is used for the Laplacian on the coarse
grids\cite{Bri96}.

When all wave functions have been updated, the wave functions are
othonormalized, and the density is updated.  From the new density, a
new Hamiltonian is generated ($\tilde{v}_G$ and $H_{i_1i_2}^a$).
Finally, a subspace diagonalization is performed, and the next
iteration towards self-consistency can begin.  For each iteration a
new input density is estimated using Pulay mixing\cite{Pul80}.
Typically three old densities are used.  The atomic density matrices,
$D_{i_1i_2}^a$, are mixed as well.  We start the iterations from a good
guess at the wave function: A linear combination of atomic orbitals.

\section{Forces}

\label{forces}
The atomic force acting on an atom is defined as
\begin{align}
  \mathbf{F}^a = & -\frac{dE}{d\mathbf{R}^a} \nonumber \\
  = & -\frac{\partial E}{\partial \mathbf{R}^a}
  - \sum_{nG} \left( \frac{\partial E}{\partial \tilde{\psi}_{nG}}
    \frac{d\tilde{\psi}_{nG}}{d\mathbf{R}^a}
    + \text{c.c.}\right) \nonumber \\
  = & -\frac{\partial E}{\partial \mathbf{R}^a} \nonumber \\
  & -V_c \sum_n f_n \epsilon_n
  \sum_{GG'} O_{GG'} \left( \tilde{\psi}_{nG'}^*
    \frac{d\tilde{\psi}_{nG}}{d\mathbf{R}^a} + \text{c.c.}\right).
\label{F}
\end{align}
In the last line, we have used Eqs.~(\ref{grad}) and (\ref{KS}).
The variation of the wave function corresponding to a
variation in the position can be found from Eq.~(\ref{O}):
\begin{equation}
  \frac{d}{d\mathbf{R}^a}
  \sum_{GG'} \tilde{\psi}_{nG}^* O_{GG'} \tilde{\psi}_{n'G'} = 0.
\end{equation}
Inserting into Eq.~(\ref{F}), we get:
\begin{align}
  \mathbf{F}^a = & -\frac{\partial E}{\partial \mathbf{R}^a}
  + V_c \sum_n f_n \epsilon_n
  \sum_{GG'} \tilde{\psi}_{nG}^* \frac{dO_{GG'}}{d\mathbf{R}^a}
  \tilde{\psi}_{nG'} \nonumber \\
  = & -V_c \sum_G \tilde{v}_G \frac{d\tilde{n}_{cG}^a}{d\mathbf{R}^a}
  - V_f \sum_g \tilde{v}_g^{\text{H}} \sum_L Q_L^a
  \frac{d\hat{g}_{Lg}^a}{d\mathbf{R}^a} \nonumber \\
  & - V_f \sum_g \tilde{n}_g (\frac{d\bar{v}_g^a}{d\mathbf{R}^a} +
  \sum_L Q_L^a \frac{d\hat{v}_{Lg}^a}{d\mathbf{R}^a}) \nonumber \\
  & - V_c \sum_n f_n \sum_{i_1i_2} 
  (H_{i_1i_2}^a - \epsilon_n O_{i_1i_2}^a) \nonumber \\
  & \quad \times \sum_G [\tilde{\psi}_{nG}
  \frac{d\tilde{p}_{i_1G}^a}{d\mathbf{R}^a}
  (P_{ni_2}^a)^* + \text{c.c.}] \nonumber \\
  & - \sum_{a'} \sum_{LL'} Q_L^a \frac{dV_{LL'}^{aa'}}{d\mathbf{R}^a}
  Q_{L'}^{a'}.
  \label{ff}
\end{align}

\begin{figure}[htbp]
  \centering
  \includegraphics[width=\linewidth,clip=]{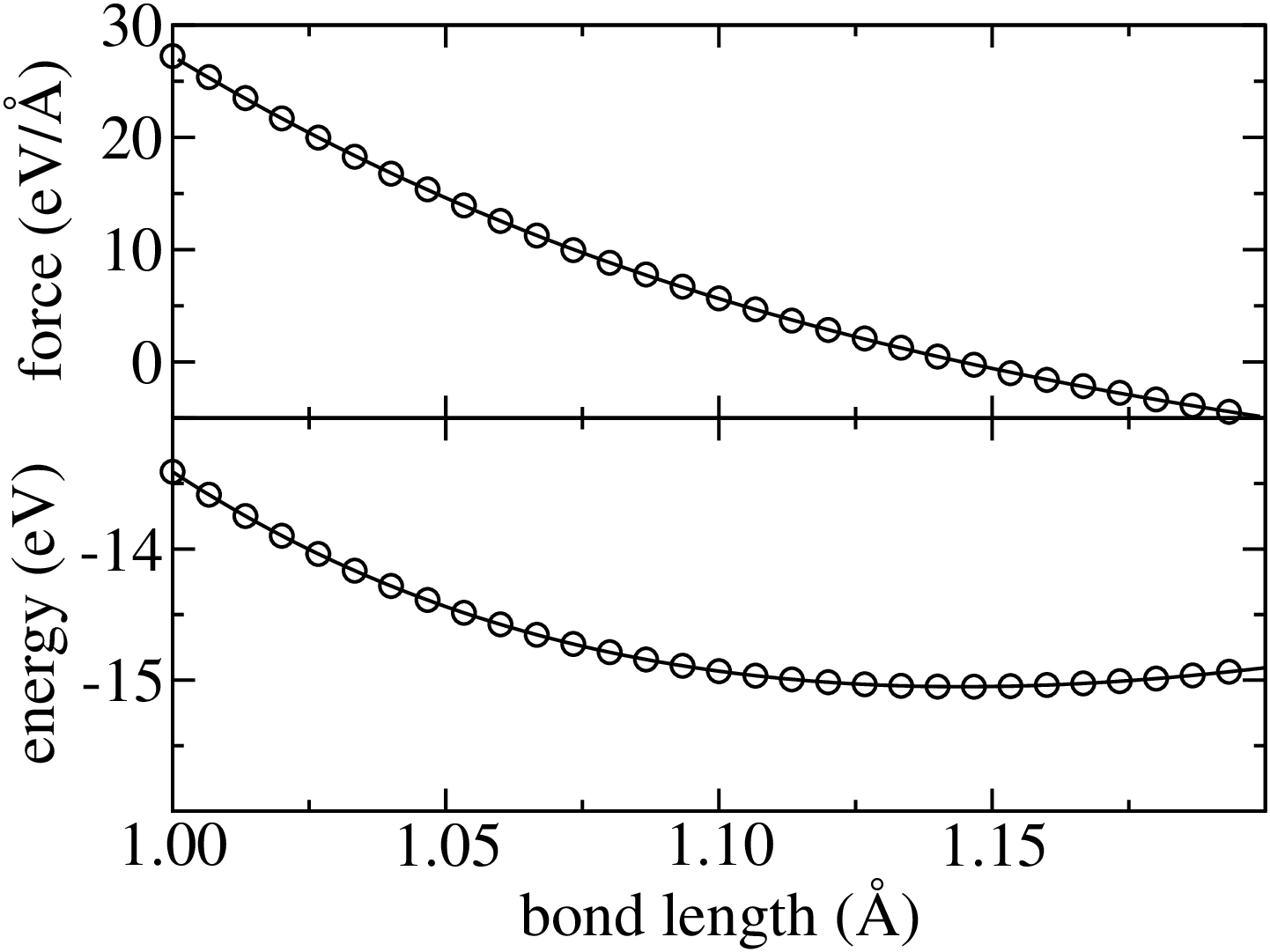}
  \caption{Energy and force for a CO molecule at different
    bond lengths calculated with $h = 0.2$ {\AA}.  Bottom: The circles
    show the calculated energies and the curve shows a third-order
    polynomial fit.  Top: The circles show the calculated forces
    according to Eq.~(\ref{ff}).  The curve is minus the derivative of the
    third order polynomial fit to the energies.}
  \label{force}
\end{figure}

As an example, we show in Fig.~\ref{force}, the force along the bond of
a CO molecule calculated using the analytical expression above.
Fitting a third order polynomial to the energies and taking the negative
of the derivative with respect to the bond length, is seen to give
exactly the same force.

\section{Generalizations}

\label{generalizations}
We have implemented the algorithms introduced above, and in addition
we have made two extensions: 1) treatment of spin-polarized systems
and 2) treatment of periodic systems using Brillouin zone sampling.
The first extension is straightforward.  When $\mathbf{k}$-points are
introduced in order to treat periodic systems, we can work directly
with the wave functions and use Bloch boundary conditions\cite{Kor96}:
\begin{equation}
  \tilde{\psi}_{n\mathbf{k}}(\mathbf{r} + \mathbf{R}) =
  e^{i \mathbf{k} \cdot \mathbf{R}} \tilde{\psi}_{n\mathbf{k}}(\mathbf{r}),
\end{equation}
where $\mathbf{R}$ is any Bravais lattice vector.  This is different from the plane wave approach, where the periodic basis set
forces one to work with the periodic part of the wave function and the
Hamiltonian becomes $\mathbf{k}$-point dependent.  In our case, the
boundary conditions become $\mathbf{k}$-point dependent.

It is only necessary to work with the $\mathbf{k}$-points in the
irreducible part of the Brillouin zone.  Each $\mathbf{k}$-point will
have a specific weight, and densities, atomic density matrices, and
forces should be appropriately symmetrized.

For evaluating the GGA exchange-correlation energy and potential, we
use a finite difference operator for calculating the gradient of the
density.  The exchange-correlation potential is calculated as the
exact derivative of the discretised exchange-correlation energy with
respect to $\tilde{n}_g$ (similar in spirit to the method of
White and Bird\cite{Whi94} used for plane-wave basis sets).

\section{Applications}

\label{applications}
The first application of the algorithms described here is the
calculation of atomization energies for the twenty small molecules
listed in table~\ref{mol20} using the PBE\cite{Per96}
exchange-correlation functional.  The augmentation sphere radii used
and the number of projectors used are shown in table~\ref{rc}.  The
choice of the augmentation sphere radii is a compromise between
smooth pseudo wave functions and a low number of projector functions:
A larger radius will allow us to have smoother pseudo wave functions,
but more projector functions will be needed for high accuracy.
Furthermore, the radius is limited by the requirement that the overlap
between neighboring augmentation spheres should be small.  The radii
we have chosen, will give slight overlaps in some of the molecular
calculations.

\begin{table}[htbp]
  \caption{PBE and experimental atomization energies (experimental
    geometries are used\cite{Per96}, and zero point vibration energy has
    been removed).  The ground states of C, O, F, P and Cl are found to be
    non-spherical\cite{Kut87}.  All-electron and experimental numbers 
    are taken from Refs.~\onlinecite{Kur99} and \onlinecite{Zha98}.}
  \begin{tabular}{lcccc}
    \hline
    \hline
    Molecule & \multicolumn{3}{c}{PBE} & Experiment\cite{Kur99} \\
             & PAW  & all-electron\cite{Kur99} & all-electron\cite{Zha98} & \\
    \hline
    H$_2$       &  4.52 &  4.54 &  4.53 &  4.75 \\
    LiH         &  2.32 &  2.32 &  2.32 &  2.51 \\
    CH$_4$      & 18.17 & 18.20 & 18.18 & 18.18 \\
    NH$_3$      & 13.03 & 13.08 & 13.05 & 12.90 \\
    OH          &  4.76 &  4.76 &  4.75 &  4.61 \\
    H$_2$O      & 10.14 & 10.16 & 10.14 & 10.07 \\
    HF          &  6.27 &  6.16 &  6.14 &  6.11 \\
    Li$_2$      &  0.89 &  0.86 &  0.85 &  1.06 \\
    LiF         &  6.16 &  6.01 &  6.05 &  6.02 \\
    Be$_2$      &  0.35 &  0.42 &  0.41 &  0.13 \\
    C$_2$H$_2$  & 17.95 & 17.99 & 17.91 & 17.58 \\
    C$_2$H$_4$  & 24.75 & 24.78 & 24.73 & 24.40 \\
    HCN         & 14.10 & 14.14 & 14.07 & 13.53 \\
    CO          & 11.58 & 11.66 & 11.60 & 11.24 \\
    N$_2$       & 10.41 & 10.55 & 10.46 &  9.91 \\
    NO          &  7.38 &  7.45 &  7.36 &  6.63 \\
    O$_2$       &  6.27 &  6.23 &  6.14 &  5.23 \\
    F$_2$       &  2.28 &  2.32 &  2.25 &  1.67 \\
    P$_2$       &  5.18 &  5.25 &  5.08 &  5.09 \\
    Cl$_2$      &  2.83 &  2.82 &  2.74 &  2.52 \\
    \hline
    \hline
  \end{tabular}
  \label{mol20}
\end{table}

\begin{table*}[htbp]
  \caption{Augmentation sphere radii in atomic units and number of projectors.}
  \begin{tabular}{lccccccccccc}
    \hline
    \hline
    atom & H & Li & Be & C & N & O & F & Al & Si & P & Cl \\
    \hline
    $r_c^a$ [Bohr] & 0.9& 1.5& 1.5& 1.0& 1.1& 1.2& 1.2& 2.0& 2.0& 2.0&
    1.5 \\
    number of projectors & s$^2$p & s$^2$p & s$^2$p & s$^2$p$^2$d  & s$^2$p$^2$d  & s$^2$p$^2$d  & s$^2$p$^2$d  & s$^2$p$^2$d  & s$^2$p$^2$d  & s$^2$p$^2$d  & s$^2$p$^2$d \\
    \hline
    \hline
  \end{tabular}
  \label{rc}
\end{table*}

For the second row atoms, the 1s orbital is treated as a core state
and frozen, and for the third row atoms, the 1s, 2s and 2p orbitals
are frozen.  For hydrogen, lithium, and beryllium, we use two
s-projectors and one p-projector, and for the rest of the atoms two
s-projectors, two p-projectors and one d-projector are used.  The
compensation charges were taken to be spherical.  We calculate the
atomic exchange-correlation correction energy, Eq.~(\ref{xca}), using
49 line integrations in each sphere\cite{Fli99}.  All calculations are
done using periodic boundary conditions.

With these approximations we get excellent agreement with full
all-electron calculations (see Table~\ref{mol20}).  The average and
maximum differences between our PAW atomization energies and the
all-electron calculations of Kurth {\it et al.}\cite{Kur99} are 0.05
eV and 0.15 eV respectively, and comparing with the all-electron
calculations of Zhang {\it et al.}\cite{Zha98}, we get an average
difference of 0.05 eV and a maximum difference of 0.13 eV (the two
sets of all-electron calculations differ by 0.05 eV in average and
0.17 as maximum).  We find that all atomization energies are converged
to within 0.03 eV/atom at a grid spacing of 0.1875 {\AA}.

\begin{figure}[htbp]
  \centering
  \includegraphics[width=\linewidth,clip=]{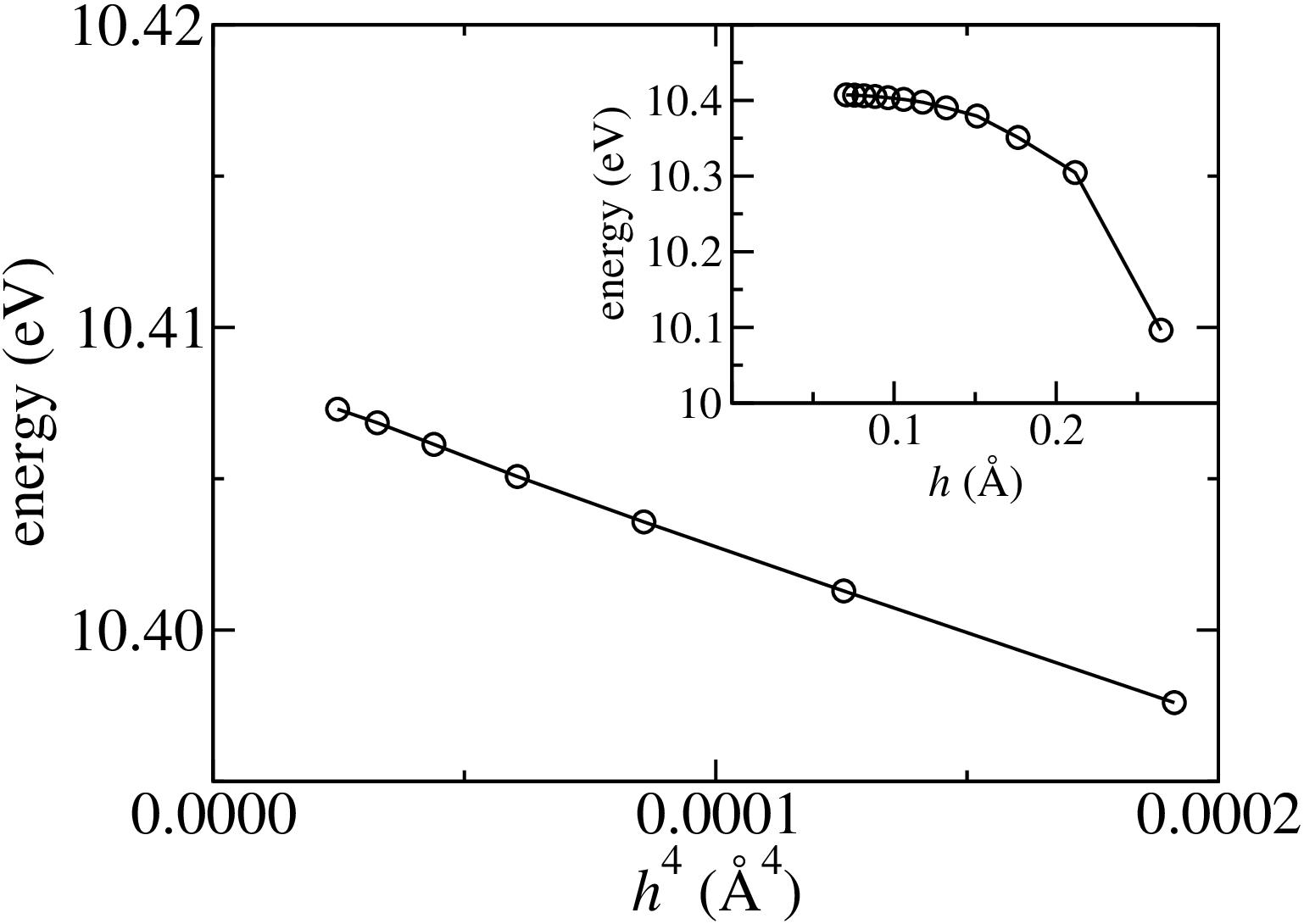}
  \caption{Atomization energy of a nitrogen molecule as a function of
    $h^4$.  The inset show the atomization energy as a function of
    $h$.}
  \label{conv}
\end{figure}

Interestingly, we find the convergence of total energies with respect
to grid spacing to be very systematic.  Figure~\ref{conv} shows the
atomization energy of nitrogen as a function of the fourth power of
the grid spacing.  It is seen that for small $h$, all points fall
exactly on a straight line, which allows us to extrapolate energies to
the limit of an infinitely dense grid ($h=0$).  The PAW numbers
presented in Table~\ref{mol20} have been extrapolated to $h=0$.  A
similarly transparent convergence of DFT calculations was recently
observed by Daykov {\it et al}.\cite{Day03} for wavelet-based
calculations.  A quartic convergence is to be expected, because all
approximations are accurate to order at least $h^3$.

\begin{figure}[htbp]
  \centering
  \includegraphics[width=\linewidth,clip=]{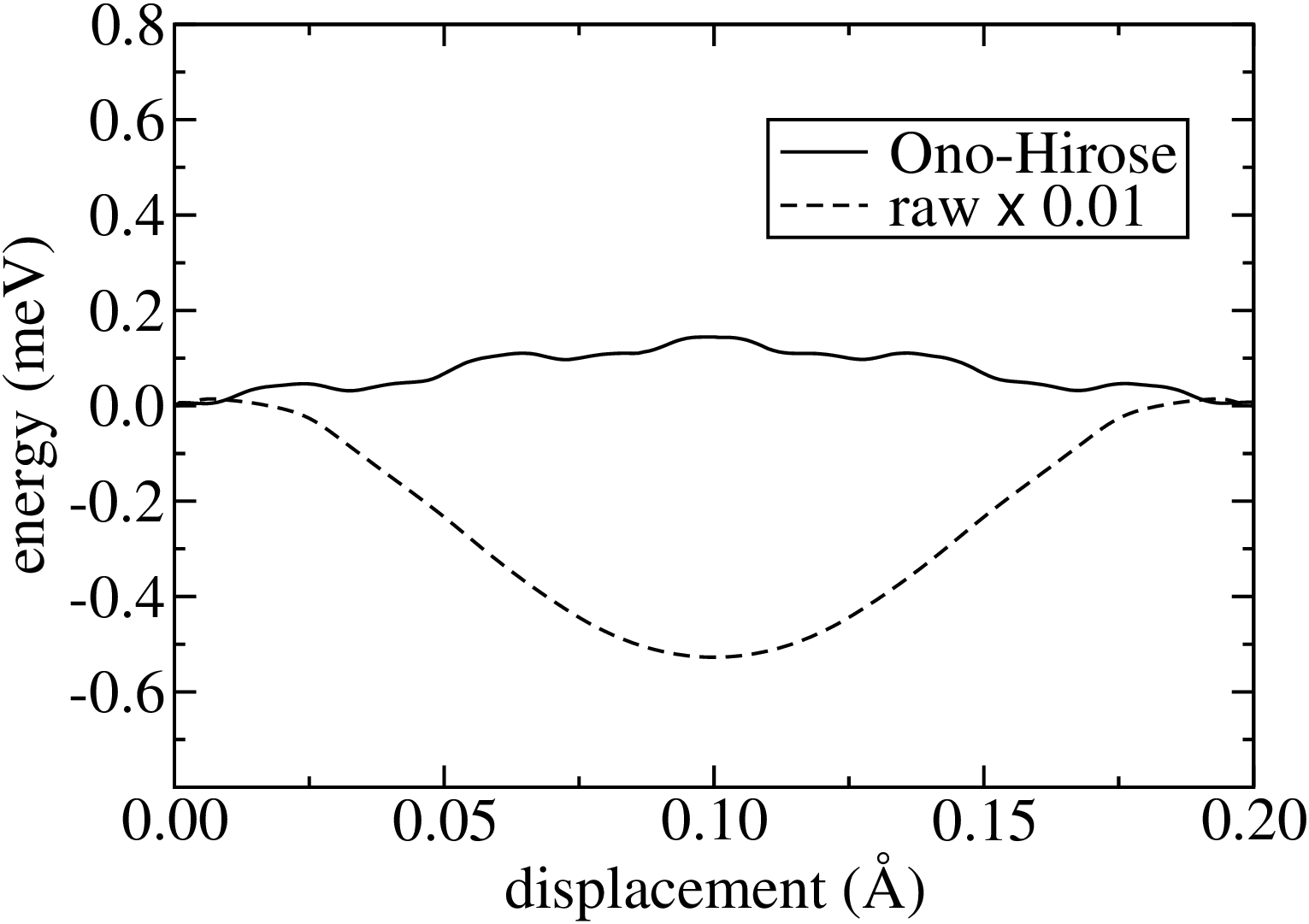}
  \caption{Energy variation as a hydrogen atom is displaced from one
  grid point to a neighboring grid point for $h = 0.2$ {\AA}.  The full
  and dashed curves show the result with and without using the
  double-grid technique (notice that the dashed curve has been
  multiplied by 0.01).}
  \label{noise}
\end{figure}

Figure \ref{noise} shows the variation of the energy as a hydrogen
atom is displaced from one grid point to a neighboring grid point.
Ideally, there should be no variation (we are using periodic boundary
conditions).  In practice, we have to make sure that this energy
variation and the corresponding forces are acceptably small.  For
hydrogen, the variation is below 0.15 meV (full line in
Fig.~\ref{noise}).  The energy varies periodically with period $h$.
Notice that there is also a modulation of the energy with a period of
$h / 5$.  This stems from the Ono-Hirose restriction of the projector
functions:  the localized projector functions are evaluated on a fine
grid with a grid spacing of $h / 5$ and then restricted to a coarse grid
with the same grid density as the wave functions.  The oscillations
give rise to forces up to 0.006 eV/{\AA} in magnitude, which is
acceptable for most applications.  The forces can be reduced further
either by using a finer grid for the wave functions, or by using a
finer grid for the Ono-Hirose restriction of the projector functions.
If the projectors are evaluated directly on the coarse grid, then the
variation of the energy is 55 meV --- more than two orders of magnitude larger
(dashed line in Fig.~\ref{noise}).  This clearly demonstrates the
importance of the Ono-Hirose restriction.

For the bulk aluminum calculation (LDA\cite{Per92} and
PBE\cite{Per96} results are shown in Table~\ref{Al}), we use a cubic unit
cell containing four aluminum atoms, and we use $10\times 10\times
10$ $\mathbf{k}$-points.  Again, we get good agreement with exact
all-electron calculations for both the lattice constant and the bulk
modulus (the bulk modulus is calculated at the theoretical lattice
constant).

\begin{table}[htbp]
  \caption{Lattice constant and Bulk modulus for fcc bulk aluminum.  All-electron and experimental numbers
    are taken from Ref.~\onlinecite{Kur99}.}
  \begin{tabular}{ccccc}
    \hline
    \hline
    XC        & \multicolumn{2}{c}{$a$ (\AA)} & \multicolumn{2}{c}{$B$ (GPa)}\\
              & PAW & all-electron             &    PAW & all-electron \\
    \hline
    LDA       & 3.987         & 3.983          & 83.6 & 84.0 \\
    PBE       & 4.043         & 4.039          & 77.7 & 77.3 \\
    \hline
    experiment& \multicolumn{2}{c}{4.050}  & \multicolumn{2}{c}{77.3}\\
    \hline
    \hline
  \end{tabular}
  \label{Al}
\end{table}


\section{Performance}

\label{performance}
We have compared the performance of the real-space code with a highly
optimized ultra-soft plane-wave code\cite{Bah02, Ham99, Dacapo}.  The
ground state of 64 Si atoms in the diamond structure was found using
both codes.  For the plane wave calculation, plane waves with kinetic
energies up to 100 eV were used and the size of the real-space grid
for fast Fourier transforms was $36\times 36\times 36$ points.  The
same grid size was used for the real-space code ($h = 0.30$ {\AA}).
After finding the electronic ground state, one atom was displaced by
0.1 {\AA}, and the time for converging to the new ground state was
measured.  The measured times were 21 and 26 minutes for the plane
wave code and real-space code respectively on a Pentium-4 2.6 GHz
Linux machine.  We have estimated the degree of convergence of the two
codes by calculating the cohesive energy of silicon.  The cohesive
energy calculated with the plane wave code and a plane wave cutoff of
100 eV, was converged to within 40 meV of the fully converged value.
With the real-space code and $h = 0.30$ {\AA}, the cohesive energy was
converged to within 3 meV of the $h = 0$ value.  This result and
similar results for other calculations we have performed seem to
indicate that the real-space code obtains a somewhat better
convergence at the same grid spacing as the plane-wave code.
Furthermore, the real-space code can, most likely, take advantage of a
number of improvements, such as algorithmic improvements and
optimizations of floating point operations and memory access.  The
plane wave code has already benefited from optimizations of this sort.

Regarding the memory requirements, the plane-wave code is clearly more
economic.  The memory required to store one wave function for the 64
atom silicon system is $36^3 = 46656$ floating point numbers for the
real-space code and 2897 floating point numbers\cite{complex} for the
plane-wave code.

\section{Discussion}

Using the techniques described in this article, an important step in
every electronic structure calculation, namely the application of the
Hamiltonian to all wave functions, can be done in $O(N^2)$ operations.
This is a more optimal scaling, than the $O(N^2\log N)$ scaling that
can be achieved with plane-wave basis sets and fast Fourier
transforms.  Operations such as orthogonalization and subspace
diagonalization of the wave functions scale as $O(N^3)$.  Luckily, the
prefactors for these $O(N^3)$ operations are very low, so that very
large system sizes are required before the $O(N^3)$ terms become the
bottleneck\cite{Tro95}.  In the limit where $O(N^3)$ terms start to
dominate, plane wave methods will in principle have an advantage,
because the number of plane wave coefficients will typically be less
than the number of grid points used in real-space grid-based
calculations.  However, we believe, that for those very large systems,
efficient parallelization on massively parallel computers and the use
of $O(N)$ methods is crucial.

Currently, we have a single processor implementation of our PAW
real-space algorithms\cite{GridPAW}.  This limits us to study rather
small systems.  Obviously a parallelization using real-space domain
decomposition is needed.  For the small systems that we have studied
so far, charge sloshing is less of a problem, but it may become a
problem when we move on to larger systems.  It will therefore be
necessary to improve on our mixing of the density.  A preconditioning,
such as that proposed by Kerker\cite{Ker81}, that will damp the long
wavelength changes to the density should be considered.  Another
improvement would be to use a special metric, that weights long
wavelength errors higher than short wavelength errors, for estimating
the norm of the difference between input and output densities in the
Pulay method\cite{Kre96}.  These improvements to the density mixing
are easily implemented in reciprocal space, but may be challenging to
do in real-space.

\section{Acknowledgements}

Financial support from the Carlsberg Foundation is gratefully
acknowledged.  We furthermore acknowledge support from the Materials
Research Program of the Danish Research Agency grant \#5020-00-0012,
and from the Danish Center for Scientific Computing.

\section{Appendix}

In the following, we provide explicit formulas needed for calculating
the constants and functions that describe an atomic species.

\subsection{Construction of partial waves and projector functions}

\label{construction}
A DFT calculation for the atom is performed using radial grids for the
wave functions, densities and potentials.  The radial Kohn-Sham
equation gives us a set of radial all-electron wave functions, normalized as
$\int r^2 dr [\phi_{n\ell}^a(r)]^2 = 1$.  The core states are only
used for constructing a frozen core electron density:
\begin{equation}
  n_c^a(r) = 2 \sum_{n\ell}^{\text{core}} \frac{2\ell + 1}{4\pi}
  [\phi_{n\ell}^{a,\text{core}}(r)]^2.
\end{equation}

The smooth partial wave functions are chosen as
\begin{equation}
  \tilde{\phi}_{n\ell}^a(r) = \sum_{i=0}^3 c_i r^{2i},
\end{equation}
for $r < r_c^a$, and the coefficients, $c_i$, are chosen so that
$\tilde{\phi}_{n\ell}^a$ joins $\phi_{n\ell}^a$ smoothly at $r =
r_c^a$.  The projector functions are calculated as
\begin{equation}
  \tilde{p}_{n\ell}^a(r) =
  (-\tfrac{1}{2}\nabla^2 + \tilde{v}(r) - \epsilon_{n\ell}^a)
 \tilde{\phi}_{n\ell}^a(r).
\end{equation}
The projector functions must be orthonormalized as described by
Bl{\"o}chl\cite{Blo94}.  The potential $\bar{v}^a$ is chosen so that
the local effective potential, $\tilde{v} = \tilde{v}^{\text{H}} +
\tilde{v}_{\text{xc}} + \bar{v}^a$, has the following shape for $r <
r_c^a$ in the atomic reference state:
\begin{equation}
  \tilde{v}(r) = a^a + b^a r^2, ~~~ r < r_c^a.
\end{equation}
The constants $a^a$ and $b^a$ are found by requiring that
$\bar{v}(r_c^a) = 0$ and $d\bar{v}(r)/dr|_{r=r_c^a} = 0$.

\subsection{Compensation charges}

For the compensation charges we use Gaussians:
\begin{equation}
  \tilde{g}^a_\ell(r) = \frac{1}{\sqrt{4\pi}}
  \frac{\ell !}{(2\ell + 1)!}
  (4\alpha^a)^{\ell + 3/2} r^\ell e^{-\alpha^a r^2}
\end{equation}
and
\begin{equation}
  \hat{g}^a_\ell(r) = \frac{1}{\sqrt{4\pi}}
  \frac{\ell !}{(2\ell + 1)!}
  (4\hat{\alpha}^a)^{\ell + 3/2} r^\ell e^{-\hat{\alpha}^a r^2}.
\end{equation}
With this choice for the compensation
charges, the integral, $V_{LL'}^{aa'}$, in Eq.~(\ref{pair}), can be
evaluated analytically\cite{Oba86}.
We choose the $\alpha$'s so that $\alpha^a(r_c^a)^2 = 9.0$ and $\hat{\alpha}^a(\hat{r}_c^a)^2 = 22.0$.

\subsection{Atomic constants}

\label{constants}
By inserting Eqs.~(\ref{n}), (\ref{nt}) and (\ref{Zt}) into
Eqs.~(\ref{ea}) and (\ref{eat}) we can reduce $E^a - \tilde{E}^a$ to:

\begin{align}
  E^a - \tilde{E}^a & = F^a
   + \sum_{i_1i_2} D_{i_1i_2}^a G_{i_1i_2}^a \nonumber \\
  & + \sum_{i_1i_2i_3i_4} D_{i_1i_2}^a I_{i_1i_2i_3i_4}^a D_{i_3i_4}^a 
  \nonumber \\
  & + \sum_L Q_L^a J_L^a \nonumber \\
  & + \sum_{i_1i_2} \sum_L D_{i_1i_2}^a Q_L^a K_{i_1i_2L}^a \nonumber \\
  & + \sum_{LL'} Q_L^a Q_{L'}^a M_{LL'}^a \nonumber \\
  & + \Delta E_{xc}^a(\{D_{i_1i_2}^a\}).
  \label{E}
\end{align}
The integrals $F^a$, $G_{i_1i_2}^a$, $I_{i_1i_2i_3i_4}^a$, $J_L^a$,
$K_{i_1i_2L}^a$ and $M_{LL'}^a$ are given below and need only be
computed once for each type of atom.

Inserting Eq.~(\ref{Q}) into Eq.~(\ref{E}) leads us to
Eq.~(\ref{ABC}), where

\begin{align}
  A^a = & F^a + \Delta^a K_{00}^a 
  + (\Delta^a)^2 N_{00,00}^a \nonumber \\   
  B_{i_1i_2}^a = & J_{i_1i_2}^a + 
  \sum_L \Delta_{Li_1i_2}^a K_L^a \nonumber \\   
  & + \Delta^a M_{i_1i_2,00}^a +
  2 \sum_{L} \Delta_{Li_1i_2}^a N_{00,00}^a \Delta^a \nonumber \\   
  C_{i_1i_2i_3i_4}^a = & J_{i_1i_2i_3i_4}^a + 
  \sum_L M_{i_1i_2L}^a \Delta_{Li_3i_4}^a \nonumber \\   
  & + \sum_{LL'} \Delta_{Li_1i_2}^a N_{LL'}^a \Delta_{L'i_3i_4}^a.
\end{align}
We use the symmetrized $C_{i_1i_2i_3i_4}^a$:
\begin{eqnarray}
  C_{i_1i_2i_3i_4}^a \leftarrow \tfrac{1}{2} 
  (C_{i_1i_2i_3i_4}^a + C_{i_3i_4i_1i_2}^a).
\end{eqnarray}

The integrals for $F^a$, $G_{i_1i_2}^a$, $I_{i_1i_2i_3i_4}^a$, $J_L^a$,
$K_{i_1i_2L}^a$ and $M_{LL'}^a$ are (all integrals are limited to
inside the augmentation spheres):
\begin{align}
  F^a = & \sum_i^{\text{core}}
  \int d \mathbf{r} \phi_i^{a,\text{core}}(\mathbf{r}) (-\tfrac{1}{2} \nabla^2)
  \phi_i^{a,\text{core}}(\mathbf{r}) \nonumber \\
  & - \int d\mathbf{r} \frac{n_c^a(r)}{r} \mathcal{Z}^a \nonumber \\
  & + \tfrac{1}{2} \int d\mathbf{r} v_c^a(r)
  [n_c^a(r) + \tilde{n}_c^a(r)] \nonumber \\
  & - \int d\mathbf{r} \tilde{n}_c^a(r) \bar{v}^a(r),
\end{align}
\begin{align}
  I_{i_1i_2}^a = & - \tfrac{1}{2}  \int d\mathbf{r} 
    [\phi_{i_1}^a(\mathbf{r}) \nabla^2 \phi_{i_2}^a(\mathbf{r}) -
    \tilde{\phi}_{i_1}^a(\mathbf{r}) \nabla^2
    \tilde{\phi}_{i_2}^a(\mathbf{r})] \nonumber \\
  & + \int d\mathbf{r}
  [v_{i_1i_2}^a(\mathbf{r})n_c(r) +
  v_c^a(r) 
  \tilde{\phi}_{i_1}^a(\mathbf{r}) \tilde{\phi}_{i_2}^a(\mathbf{r})] 
  \nonumber \\
  & - \int d\mathbf{r}
  \frac{\phi_{i_1}^a(\mathbf{r}) \phi_{i_2}^a(\mathbf{r})}{r} \mathcal{Z}^a
  \nonumber \\
  & - \int d\mathbf{r}
  \tilde{\phi}_{i_1}^a(\mathbf{r})
  \tilde{\phi}_{i_2}^a(\mathbf{r}) \bar{v}^a(r),
\end{align}
\begin{equation}
  J_{i_1i_2i_3i_4}^a =
  \tfrac{1}{2}
  \int d\mathbf{r} v_{i_1i_2}^a(\mathbf{r})
  [\phi_{i_3}^a(\mathbf{r}) \phi_{i_4}^a(\mathbf{r})
  + \tilde{\phi}_{i_3}^a(\mathbf{r})
  \tilde{\phi}_{i_4}^a(\mathbf{r})],
\end{equation}
\begin{equation}
  K_L^a = - \int d\mathbf{r} \int d\mathbf{r}'
  \frac{\tilde{g}_L^a(\mathbf{r}')}{|\mathbf{r} - \mathbf{r}'|}
  \tilde{n}_c^a(r),
\end{equation}
\begin{equation}
  M_{i_1i_2L}^a = -\int d\mathbf{r} \int d\mathbf{r}'
  \frac{\tilde{g}_L^a(\mathbf{r}')}{|\mathbf{r} - \mathbf{r}'|}
  \tilde{\phi}_{i_1}^a(\mathbf{r}) \tilde{\phi}_{i_2}^a(\mathbf{r})
\end{equation}
and
\begin{equation}
  N_{LL'}^a = -\tfrac{1}{2}
  \int d\mathbf{r} \int d\mathbf{r}'
  \frac{\tilde{g}_L^a(\mathbf{r}')}{|\mathbf{r} - \mathbf{r}'|}
  \tilde{g}_{L'}^a(\mathbf{r}).
\end{equation}

The potentials $v_{i_1i_2}^a(\mathbf{r})$ and $v_c^a(r)$ are defined as
\begin{equation}
  v_{i_1i_2}^a(\mathbf{r}) = \int d\mathbf{r'}
  \frac{\phi_{i_1}^a(\mathbf{r}') \phi_{i_2}^a(\mathbf{r}')
    - \tilde{\phi}_{i_1}^a(\mathbf{r}') \tilde{\phi}_{i_2}^a(\mathbf{r'})}
  {|\mathbf{r} - \mathbf{r}'|}
\end{equation}
and 
\begin{equation}
  v_c^a(r) = \int d\mathbf{r'} \frac{n_c^a(r') - \tilde{n}_c^a(r')}
  {|\mathbf{r} - \mathbf{r}'|}.
\end{equation}

It is advantageous to decompose $v_{i_1i_2}^a(\mathbf{r})$ into
angular momentum contributions as
\begin{equation}
  v_{i_1i_2}^a(\mathbf{r}) = \sum_L
  v_{i_1i_2\ell}^a(r) Y_L(\mathbf{r}),
\end{equation}
and solve
\begin{eqnarray}
  & \nabla^2[v_{i_1i_2\ell}^a(r) Y_L(\hat{\mathbf{r}})] = -4\pi
  G_{L_1L_2}^L Y_L(\hat{\mathbf{r}}) \nonumber \\
  & \times [\phi_{n_1\ell_1}^a(r) \phi_{n_2\ell_2}^a(r)
  - \tilde{\phi}_{n_1\ell_1}^a(r) \tilde{\phi}_{n_2\ell_2}^a(r)],
\end{eqnarray}
where
\begin{equation}
  G_{L_1L_2}^L = \int_0^\pi \sin\theta d\theta \int_0^{2\pi} d\phi
  Y_L(\theta, \phi) Y_{L_1}(\theta, \phi) Y_{L_2}(\theta, \phi)
\end{equation}
is a Gaunt coefficient.

\subsection{Hamiltonian}

\label{ham}
The discretized Hamiltonian, Eq.~(\ref{H}), depends on $\tilde{v}_G$ and
$H^a_{i_1i_2}$.  The local effective potential on the coarse
grid, $\tilde{v}_G = (V_f / V_c) \sum_g I_{gG} \tilde{v}_g$, is a
restriction of the local effective potential on the fine grid:
\begin{equation}
  \tilde{v}_g = 
  \tilde{v}_g^{\text{H}}
  + \sum_a \hat{v}_g^a + \sum_a \bar{v}_g^a 
  + \frac{1}{V_f} \frac{\partial E_{\text{xc}}}{\partial \tilde{n}_g},
  \label{vt_g}
\end{equation}
and the atomic Hamiltonian, $H^a_{i_1i_2}$, is
\begin{align}
  H^a_{i_1i_2} = & \sum_L\Delta^a_{Li_1i_2} W_L^a
  + \frac{\partial \Delta E_{xc}^a}{\partial D_{i_1i_2}^a}
  \nonumber \\
  & + B_{i_1i_2}^a
  + 2 \sum_{i_3i_4} C_{i_1i_2i_3i_4}^a D_{i_3i_4}^a,
\end{align}
where
\begin{align}
  W_L^a & = \frac{\partial \tilde{E}}{\partial Q_L^a} \nonumber \\
  & = V_f \sum_g \tilde{v}^{\text{H}}_g \hat{g}_{Lg}^a
  + V_f \sum_g \tilde{n}_g \hat{v}_{Lg}^a \nonumber \\
  & + \sum_{a'}\sum_{L'} V_{LL'}^{aa'} Q_{L'}^{a'}.
\end{align}

\bibliographystyle{apsrev}

\begin{thebibliography}{57}
\expandafter\ifx\csname natexlab\endcsname\relax\def\natexlab#1{#1}\fi
\expandafter\ifx\csname bibnamefont\endcsname\relax
  \def\bibnamefont#1{#1}\fi
\expandafter\ifx\csname bibfnamefont\endcsname\relax
  \def\bibfnamefont#1{#1}\fi
\expandafter\ifx\csname citenamefont\endcsname\relax
  \def\citenamefont#1{#1}\fi
\expandafter\ifx\csname url\endcsname\relax
  \def\url#1{\texttt{#1}}\fi
\expandafter\ifx\csname urlprefix\endcsname\relax\def\urlprefix{URL }\fi
\providecommand{\bibinfo}[2]{#2}
\providecommand{\eprint}[2][]{\url{#2}}

\bibitem[{\citenamefont{Hohenberg and Kohn}(1964)}]{Hoh64}
\bibinfo{author}{\bibfnamefont{P.}~\bibnamefont{Hohenberg}} \bibnamefont{and}
  \bibinfo{author}{\bibfnamefont{W.}~\bibnamefont{Kohn}},
  \bibinfo{journal}{Phys. Rev.} \textbf{\bibinfo{volume}{136}},
  \bibinfo{pages}{B864} (\bibinfo{year}{1964}).

\bibitem[{\citenamefont{Kohn and Sham}(1965)}]{Koh65}
\bibinfo{author}{\bibfnamefont{W.}~\bibnamefont{Kohn}} \bibnamefont{and}
  \bibinfo{author}{\bibfnamefont{L.~J.} \bibnamefont{Sham}},
  \bibinfo{journal}{Phys. Rev.} \textbf{\bibinfo{volume}{140}},
  \bibinfo{pages}{A1133} (\bibinfo{year}{1965}).

\bibitem[{\citenamefont{Chelikowsky et~al.}(1994)\citenamefont{Chelikowsky,
  Troullier, Wu, and Saad}}]{Che94}
\bibinfo{author}{\bibfnamefont{J.~R.} \bibnamefont{Chelikowsky}},
  \bibinfo{author}{\bibfnamefont{N.}~\bibnamefont{Troullier}},
  \bibinfo{author}{\bibfnamefont{K.}~\bibnamefont{Wu}}, \bibnamefont{and}
  \bibinfo{author}{\bibfnamefont{Y.}~\bibnamefont{Saad}},
  \bibinfo{journal}{Phys. Rev. B} \textbf{\bibinfo{volume}{50}},
  \bibinfo{pages}{11355} (\bibinfo{year}{1994}).

\bibitem[{\citenamefont{Seitsonen et~al.}(1995)\citenamefont{Seitsonen, Puska,
  and Nieminen}}]{Sei95}
\bibinfo{author}{\bibfnamefont{A.~P.} \bibnamefont{Seitsonen}},
  \bibinfo{author}{\bibfnamefont{M.~J.} \bibnamefont{Puska}}, \bibnamefont{and}
  \bibinfo{author}{\bibfnamefont{R.~M.} \bibnamefont{Nieminen}},
  \bibinfo{journal}{Phys. Rev. B} \textbf{\bibinfo{volume}{51}},
  \bibinfo{pages}{14057} (\bibinfo{year}{1995}).

\bibitem[{\citenamefont{Hoshi et~al.}(1995)\citenamefont{Hoshi, Arai, and
  Fujiwara}}]{Hos95}
\bibinfo{author}{\bibfnamefont{T.}~\bibnamefont{Hoshi}},
  \bibinfo{author}{\bibfnamefont{M.}~\bibnamefont{Arai}}, \bibnamefont{and}
  \bibinfo{author}{\bibfnamefont{T.}~\bibnamefont{Fujiwara}},
  \bibinfo{journal}{Phys. Rev. B} \textbf{\bibinfo{volume}{52}},
  \bibinfo{pages}{5459} (\bibinfo{year}{1995}).

\bibitem[{\citenamefont{Briggs et~al.}(1996)\citenamefont{Briggs, Sullivan, and
  Bernholc}}]{Bri96}
\bibinfo{author}{\bibfnamefont{E.~L.} \bibnamefont{Briggs}},
  \bibinfo{author}{\bibfnamefont{D.~J.} \bibnamefont{Sullivan}},
  \bibnamefont{and} \bibinfo{author}{\bibfnamefont{J.}~\bibnamefont{Bernholc}},
  \bibinfo{journal}{Phys. Rev. B} \textbf{\bibinfo{volume}{54}},
  \bibinfo{pages}{14362} (\bibinfo{year}{1996}).

\bibitem[{\citenamefont{Bernholc et~al.}(1997)\citenamefont{Bernholc, Briggs,
  Sullivan, Brabec, Nardelli, Rapcewicz, Roland, and Wensell}}]{Ber97}
\bibinfo{author}{\bibfnamefont{J.}~\bibnamefont{Bernholc}},
  \bibinfo{author}{\bibfnamefont{E.~L.} \bibnamefont{Briggs}},
  \bibinfo{author}{\bibfnamefont{D.~J.} \bibnamefont{Sullivan}},
  \bibinfo{author}{\bibfnamefont{C.~J.} \bibnamefont{Brabec}},
  \bibinfo{author}{\bibfnamefont{M.~B.} \bibnamefont{Nardelli}},
  \bibinfo{author}{\bibfnamefont{K.}~\bibnamefont{Rapcewicz}},
  \bibinfo{author}{\bibfnamefont{C.}~\bibnamefont{Roland}}, \bibnamefont{and}
  \bibinfo{author}{\bibfnamefont{M.}~\bibnamefont{Wensell}},
  \bibinfo{journal}{Int. J. Quantum Chem.} \textbf{\bibinfo{volume}{65}},
  \bibinfo{pages}{531} (\bibinfo{year}{1997}).

\bibitem[{\citenamefont{Ancilotto et~al.}(1999)\citenamefont{Ancilotto,
  Blandin, and Toigo}}]{Anc99}
\bibinfo{author}{\bibfnamefont{F.}~\bibnamefont{Ancilotto}},
  \bibinfo{author}{\bibfnamefont{P.}~\bibnamefont{Blandin}}, \bibnamefont{and}
  \bibinfo{author}{\bibfnamefont{F.}~\bibnamefont{Toigo}},
  \bibinfo{journal}{Phys. Rev. B} \textbf{\bibinfo{volume}{59}},
  \bibinfo{pages}{7868} (\bibinfo{year}{1999}).

\bibitem[{\citenamefont{Bernholc et~al.}(2000)\citenamefont{Bernholc, Briggs,
  Bungaro, Nardelli, Fattebert, Rapcewicz, Roland, Schmidt, and Zhao}}]{Ber00}
\bibinfo{author}{\bibfnamefont{J.}~\bibnamefont{Bernholc}},
  \bibinfo{author}{\bibfnamefont{E.~L.} \bibnamefont{Briggs}},
  \bibinfo{author}{\bibfnamefont{C.}~\bibnamefont{Bungaro}},
  \bibinfo{author}{\bibfnamefont{M.~B.} \bibnamefont{Nardelli}},
  \bibinfo{author}{\bibfnamefont{J.-L.} \bibnamefont{Fattebert}},
  \bibinfo{author}{\bibfnamefont{K.}~\bibnamefont{Rapcewicz}},
  \bibinfo{author}{\bibfnamefont{C.}~\bibnamefont{Roland}},
  \bibinfo{author}{\bibfnamefont{W.~G.} \bibnamefont{Schmidt}},
  \bibnamefont{and} \bibinfo{author}{\bibfnamefont{Q.}~\bibnamefont{Zhao}},
  \bibinfo{journal}{Phys. Stat. Sol.} \textbf{\bibinfo{volume}{217}},
  \bibinfo{pages}{685} (\bibinfo{year}{2000}).

\bibitem[{\citenamefont{Beck}(2000)}]{Bec00}
\bibinfo{author}{\bibfnamefont{T.~L.} \bibnamefont{Beck}},
  \bibinfo{journal}{Rev. Mod. Phys.} \textbf{\bibinfo{volume}{72}},
  \bibinfo{pages}{1041} (\bibinfo{year}{2000}).

\bibitem[{\citenamefont{Wang and Beck}(2000)}]{Wan00}
\bibinfo{author}{\bibfnamefont{J.}~\bibnamefont{Wang}} \bibnamefont{and}
  \bibinfo{author}{\bibfnamefont{T.~L.} \bibnamefont{Beck}},
  \bibinfo{journal}{J. Chem. Phys.} \textbf{\bibinfo{volume}{112}},
  \bibinfo{pages}{9223} (\bibinfo{year}{2000}).

\bibitem[{\citenamefont{Heiskanen et~al.}(2001)\citenamefont{Heiskanen, Torsti,
  Puska, and Nieminen}}]{Hei01}
\bibinfo{author}{\bibfnamefont{M.}~\bibnamefont{Heiskanen}},
  \bibinfo{author}{\bibfnamefont{T.}~\bibnamefont{Torsti}},
  \bibinfo{author}{\bibfnamefont{M.~J.} \bibnamefont{Puska}}, \bibnamefont{and}
  \bibinfo{author}{\bibfnamefont{R.~M.} \bibnamefont{Nieminen}},
  \bibinfo{journal}{Phys. Rev. B} \textbf{\bibinfo{volume}{63}},
  \bibinfo{pages}{245106} (\bibinfo{year}{2001}).

\bibitem[{\citenamefont{Waghmare et~al.}(2001)\citenamefont{Waghmare, Kim,
  Park, Modine, Maragakis, and Kaxiras}}]{Wag01}
\bibinfo{author}{\bibfnamefont{U.~V.} \bibnamefont{Waghmare}},
  \bibinfo{author}{\bibfnamefont{H.}~\bibnamefont{Kim}},
  \bibinfo{author}{\bibfnamefont{I.~J.} \bibnamefont{Park}},
  \bibinfo{author}{\bibfnamefont{N.}~\bibnamefont{Modine}},
  \bibinfo{author}{\bibfnamefont{P.}~\bibnamefont{Maragakis}},
  \bibnamefont{and} \bibinfo{author}{\bibfnamefont{E.}~\bibnamefont{Kaxiras}},
  \bibinfo{journal}{Comp. Phys. Comm.} \textbf{\bibinfo{volume}{137}},
  \bibinfo{pages}{341} (\bibinfo{year}{2001}).

\bibitem[{\citenamefont{Takahashi et~al.}(2001)\citenamefont{Takahashi, Hori,
  Wakabayashi, and Nitta}}]{Tak01}
\bibinfo{author}{\bibfnamefont{H.}~\bibnamefont{Takahashi}},
  \bibinfo{author}{\bibfnamefont{T.}~\bibnamefont{Hori}},
  \bibinfo{author}{\bibfnamefont{T.}~\bibnamefont{Wakabayashi}},
  \bibnamefont{and} \bibinfo{author}{\bibfnamefont{T.}~\bibnamefont{Nitta}},
  \bibinfo{journal}{J. Phys. Chem. A} \textbf{\bibinfo{volume}{105}},
  \bibinfo{pages}{4351} (\bibinfo{year}{2001}).

\bibitem[{\citenamefont{Marques et~al.}(2003)\citenamefont{Marques, Castro,
  Bertsch, and Rubio}}]{Mar03}
\bibinfo{author}{\bibfnamefont{M.~A.~L.} \bibnamefont{Marques}},
  \bibinfo{author}{\bibfnamefont{A.}~\bibnamefont{Castro}},
  \bibinfo{author}{\bibfnamefont{G.~F.} \bibnamefont{Bertsch}},
  \bibnamefont{and} \bibinfo{author}{\bibfnamefont{A.}~\bibnamefont{Rubio}},
  \bibinfo{journal}{Computer Physics Communications}
  \textbf{\bibinfo{volume}{151}}, \bibinfo{pages}{60} (\bibinfo{year}{2003}).

\bibitem[{\citenamefont{Schmid}(2004)}]{Sch04}
\bibinfo{author}{\bibfnamefont{R.}~\bibnamefont{Schmid}}, \bibinfo{journal}{J.
  Comput. Chem.} \textbf{\bibinfo{volume}{25}}, \bibinfo{pages}{799}
  (\bibinfo{year}{2004}).

\bibitem[{\citenamefont{Shimojo et~al.}(2001)\citenamefont{Shimojo, Kalia,
  Nakano, and Vashishta}}]{Shi01}
\bibinfo{author}{\bibfnamefont{F.}~\bibnamefont{Shimojo}},
  \bibinfo{author}{\bibfnamefont{R.~K.} \bibnamefont{Kalia}},
  \bibinfo{author}{\bibfnamefont{A.}~\bibnamefont{Nakano}}, \bibnamefont{and}
  \bibinfo{author}{\bibfnamefont{P.}~\bibnamefont{Vashishta}},
  \bibinfo{journal}{Comp. Phys. Comm.} \textbf{\bibinfo{volume}{140}},
  \bibinfo{pages}{303} (\bibinfo{year}{2001}).

\bibitem[{\citenamefont{Liu et~al.}(2003)\citenamefont{Liu, Yarne, and
  Tuckerman}}]{Liu03}
\bibinfo{author}{\bibfnamefont{Y.}~\bibnamefont{Liu}},
  \bibinfo{author}{\bibfnamefont{D.~A.} \bibnamefont{Yarne}}, \bibnamefont{and}
  \bibinfo{author}{\bibfnamefont{M.~E.} \bibnamefont{Tuckerman}},
  \bibinfo{journal}{Phys. Rev. B} \textbf{\bibinfo{volume}{68}},
  \bibinfo{pages}{125110} (\bibinfo{year}{2003}).

\bibitem[{\citenamefont{Brandt}(1977)}]{Bra77}
\bibinfo{author}{\bibfnamefont{A.}~\bibnamefont{Brandt}},
  \bibinfo{journal}{Math. Comput.} \textbf{\bibinfo{volume}{31}},
  \bibinfo{pages}{333} (\bibinfo{year}{1977}).

\bibitem[{\citenamefont{Mostofi et~al.}(2002)\citenamefont{Mostofi, Skylaris,
  Haynes, and Payne}}]{Mos02}
\bibinfo{author}{\bibfnamefont{A.~A.} \bibnamefont{Mostofi}},
  \bibinfo{author}{\bibfnamefont{C.-K.} \bibnamefont{Skylaris}},
  \bibinfo{author}{\bibfnamefont{P.~D.} \bibnamefont{Haynes}},
  \bibnamefont{and} \bibinfo{author}{\bibfnamefont{M.~C.} \bibnamefont{Payne}},
  \bibinfo{journal}{Comp. Phys. Comm.} \textbf{\bibinfo{volume}{147}},
  \bibinfo{pages}{788} (\bibinfo{year}{2002}).

\bibitem[{\citenamefont{Galli and Parrinello}(1992)}]{Gal92}
\bibinfo{author}{\bibfnamefont{G.}~\bibnamefont{Galli}} \bibnamefont{and}
  \bibinfo{author}{\bibfnamefont{M.}~\bibnamefont{Parrinello}},
  \bibinfo{journal}{Phys. Rev. Lett.} \textbf{\bibinfo{volume}{69}},
  \bibinfo{pages}{3547} (\bibinfo{year}{1992}).

\bibitem[{\citenamefont{Goedecker et~al.}(2003)\citenamefont{Goedecker, Boulet,
  and Deutsch}}]{Goe03}
\bibinfo{author}{\bibfnamefont{S.}~\bibnamefont{Goedecker}},
  \bibinfo{author}{\bibfnamefont{M.}~\bibnamefont{Boulet}}, \bibnamefont{and}
  \bibinfo{author}{\bibfnamefont{T.}~\bibnamefont{Deutsch}},
  \bibinfo{journal}{Comput. Phys. Comm.} \textbf{\bibinfo{volume}{154}},
  \bibinfo{pages}{105} (\bibinfo{year}{2003}).

\bibitem[{\citenamefont{Bl{\"o}chl}(1995)}]{Blo95}
\bibinfo{author}{\bibfnamefont{P.~E.} \bibnamefont{Bl{\"o}chl}},
  \bibinfo{journal}{J. Chem. Phys.} \textbf{\bibinfo{volume}{103}},
  \bibinfo{pages}{7422} (\bibinfo{year}{1995}).

\bibitem[{\citenamefont{Martyna and Tuckerman}(1999)}]{Mar99}
\bibinfo{author}{\bibfnamefont{G.~J.} \bibnamefont{Martyna}} \bibnamefont{and}
  \bibinfo{author}{\bibfnamefont{M.~E.} \bibnamefont{Tuckerman}},
  \bibinfo{journal}{J. Chem. Phys.} \textbf{\bibinfo{volume}{110}},
  \bibinfo{pages}{2810} (\bibinfo{year}{1999}).

\bibitem[{\citenamefont{Vanderbilt}(1990)}]{Van90}
\bibinfo{author}{\bibfnamefont{D.}~\bibnamefont{Vanderbilt}},
  \bibinfo{journal}{Phys. Rev. B} \textbf{\bibinfo{volume}{41}},
  \bibinfo{pages}{7892} (\bibinfo{year}{1990}).

\bibitem[{\citenamefont{Laasonen et~al.}(1992)\citenamefont{Laasonen,
  Pasquarello, Car, Lee, and Vanderbilt}}]{Laa92}
\bibinfo{author}{\bibfnamefont{K.}~\bibnamefont{Laasonen}},
  \bibinfo{author}{\bibfnamefont{A.}~\bibnamefont{Pasquarello}},
  \bibinfo{author}{\bibfnamefont{R.}~\bibnamefont{Car}},
  \bibinfo{author}{\bibfnamefont{C.}~\bibnamefont{Lee}}, \bibnamefont{and}
  \bibinfo{author}{\bibfnamefont{D.}~\bibnamefont{Vanderbilt}},
  \bibinfo{journal}{Phys. Rev. B} \textbf{\bibinfo{volume}{47}},
  \bibinfo{pages}{10142} (\bibinfo{year}{1992}).

\bibitem[{\citenamefont{Bl{\"o}chl}(1994)}]{Blo94}
\bibinfo{author}{\bibfnamefont{P.~E.} \bibnamefont{Bl{\"o}chl}},
  \bibinfo{journal}{Phys. Rev. B} \textbf{\bibinfo{volume}{50}},
  \bibinfo{pages}{17953} (\bibinfo{year}{1994}).

\bibitem[{\citenamefont{Bl{\"o}chl et~al.}(2003)\citenamefont{Bl{\"o}chl,
  Forst, and Schimpl}}]{Blo03}
\bibinfo{author}{\bibfnamefont{P.~E.} \bibnamefont{Bl{\"o}chl}},
  \bibinfo{author}{\bibfnamefont{C.~J.} \bibnamefont{Forst}}, \bibnamefont{and}
  \bibinfo{author}{\bibfnamefont{J.}~\bibnamefont{Schimpl}},
  \bibinfo{journal}{Bulletin of Materials Science}
  \textbf{\bibinfo{volume}{26}}, \bibinfo{pages}{33} (\bibinfo{year}{2003}).

\bibitem[{\citenamefont{Holzwarth et~al.}(1997)\citenamefont{Holzwarth,
  Matthews, Dunning, Tackett, and Zeng}}]{Hol97}
\bibinfo{author}{\bibfnamefont{N.~A.~W.} \bibnamefont{Holzwarth}},
  \bibinfo{author}{\bibfnamefont{G.~E.} \bibnamefont{Matthews}},
  \bibinfo{author}{\bibfnamefont{R.~B.} \bibnamefont{Dunning}},
  \bibinfo{author}{\bibfnamefont{A.~R.} \bibnamefont{Tackett}},
  \bibnamefont{and} \bibinfo{author}{\bibfnamefont{Y.}~\bibnamefont{Zeng}},
  \bibinfo{journal}{Phys. Rev. B} \textbf{\bibinfo{volume}{55}},
  \bibinfo{pages}{2005} (\bibinfo{year}{1997}).

\bibitem[{\citenamefont{Kresse and Joubert}(1999)}]{Kre99}
\bibinfo{author}{\bibfnamefont{G.}~\bibnamefont{Kresse}} \bibnamefont{and}
  \bibinfo{author}{\bibfnamefont{D.}~\bibnamefont{Joubert}},
  \bibinfo{journal}{Phys. Rev. B} \textbf{\bibinfo{volume}{59}},
  \bibinfo{pages}{1758} (\bibinfo{year}{1999}).

\bibitem[{\citenamefont{Valiev and Weare}(1999)}]{Val99}
\bibinfo{author}{\bibfnamefont{M.}~\bibnamefont{Valiev}} \bibnamefont{and}
  \bibinfo{author}{\bibfnamefont{J.~H.} \bibnamefont{Weare}},
  \bibinfo{journal}{J. Phys. Chem.} \textbf{\bibinfo{volume}{103}},
  \bibinfo{pages}{10588} (\bibinfo{year}{1999}).

\bibitem[{\citenamefont{Tackett et~al.}(2001)\citenamefont{Tackett, Holzwarth,
  and Matthews}}]{Tac01}
\bibinfo{author}{\bibfnamefont{A.~R.} \bibnamefont{Tackett}},
  \bibinfo{author}{\bibfnamefont{N.~A.~W.} \bibnamefont{Holzwarth}},
  \bibnamefont{and} \bibinfo{author}{\bibfnamefont{G.~E.}
  \bibnamefont{Matthews}}, \bibinfo{journal}{Comput. Phys. Comm.}
  \textbf{\bibinfo{volume}{135}}, \bibinfo{pages}{348} (\bibinfo{year}{2001}).

\bibitem[{\citenamefont{Kresse and Furthmuller}(1996)}]{Kre96}
\bibinfo{author}{\bibfnamefont{G.}~\bibnamefont{Kresse}} \bibnamefont{and}
  \bibinfo{author}{\bibfnamefont{J.}~\bibnamefont{Furthmuller}},
  \bibinfo{journal}{Phys. Rev. B} \textbf{\bibinfo{volume}{54}},
  \bibinfo{pages}{11169} (\bibinfo{year}{1996}).

\bibitem[{\citenamefont{Pulay}(1980)}]{Pul80}
\bibinfo{author}{\bibfnamefont{P.}~\bibnamefont{Pulay}},
  \bibinfo{journal}{Chem. Phys. Lett.} \textbf{\bibinfo{volume}{73}},
  \bibinfo{pages}{393} (\bibinfo{year}{1980}).

\bibitem[{\citenamefont{Wood and Zunger}(1985)}]{Woo85}
\bibinfo{author}{\bibfnamefont{D.~M.} \bibnamefont{Wood}} \bibnamefont{and}
  \bibinfo{author}{\bibfnamefont{A.}~\bibnamefont{Zunger}},
  \bibinfo{journal}{J. Phys. A} \textbf{\bibinfo{volume}{18}},
  \bibinfo{pages}{1343} (\bibinfo{year}{1985}).

\bibitem[{\citenamefont{Fliege and Maier}(1999)}]{Fli99}
\bibinfo{author}{\bibfnamefont{J.}~\bibnamefont{Fliege}} \bibnamefont{and}
  \bibinfo{author}{\bibfnamefont{U.}~\bibnamefont{Maier}},
  \bibinfo{journal}{IMA Journal of Numerical Analysis}
  \textbf{\bibinfo{volume}{19}}, \bibinfo{pages}{317} (\bibinfo{year}{1999}),
  \bibinfo{note}{http://www.mathematik.uni-dortmund.de/lsx/research/projects/f%
liege/nodes/nodes.html}.

\bibitem[{sim()}]{simple}
\bibinfo{note}{Although we find this method to be the simplest for GGA
  functionals, this is probably the most complicated part of the code.}

\bibitem[{\citenamefont{Zheng and Alml{\"o}f}(1996)}]{Zhe96}
\bibinfo{author}{\bibfnamefont{Y.~C.} \bibnamefont{Zheng}} \bibnamefont{and}
  \bibinfo{author}{\bibfnamefont{J.~E.} \bibnamefont{Alml{\"o}f}},
  \bibinfo{journal}{J. Mol. Structure} \textbf{\bibinfo{volume}{388}},
  \bibinfo{pages}{277} (\bibinfo{year}{1996}).

\bibitem[{\citenamefont{Berghold et~al.}(1998)\citenamefont{Berghold, Hutter,
  and Parrinello}}]{Ber98}
\bibinfo{author}{\bibfnamefont{C.}~\bibnamefont{Berghold}},
  \bibinfo{author}{\bibfnamefont{J.}~\bibnamefont{Hutter}}, \bibnamefont{and}
  \bibinfo{author}{\bibfnamefont{M.}~\bibnamefont{Parrinello}},
  \bibinfo{journal}{Theor. Chem. Acc.} \textbf{\bibinfo{volume}{99}},
  \bibinfo{pages}{344} (\bibinfo{year}{1998}).

\bibitem[{\citenamefont{Ono and Hirose}(1999)}]{Ono99}
\bibinfo{author}{\bibfnamefont{T.}~\bibnamefont{Ono}} \bibnamefont{and}
  \bibinfo{author}{\bibfnamefont{K.}~\bibnamefont{Hirose}},
  \bibinfo{journal}{Phys. Rev. Lett.} \textbf{\bibinfo{volume}{82}},
  \bibinfo{pages}{5016} (\bibinfo{year}{1999}).

\bibitem[{\citenamefont{Payne et~al.}(1992)\citenamefont{Payne, Teter, Allan,
  Arias, and Joannopoulos}}]{Pay92}
\bibinfo{author}{\bibfnamefont{M.~C.} \bibnamefont{Payne}},
  \bibinfo{author}{\bibfnamefont{M.~P.} \bibnamefont{Teter}},
  \bibinfo{author}{\bibfnamefont{D.~C.} \bibnamefont{Allan}},
  \bibinfo{author}{\bibfnamefont{T.~A.} \bibnamefont{Arias}}, \bibnamefont{and}
  \bibinfo{author}{\bibfnamefont{J.~D.} \bibnamefont{Joannopoulos}},
  \bibinfo{journal}{Rev. Mod. Phys.} \textbf{\bibinfo{volume}{64}},
  \bibinfo{pages}{1045} (\bibinfo{year}{1992}).

\bibitem[{\citenamefont{Korhonen et~al.}(1996)\citenamefont{Korhonen, Puska,
  and Nieminen}}]{Kor96}
\bibinfo{author}{\bibfnamefont{T.}~\bibnamefont{Korhonen}},
  \bibinfo{author}{\bibfnamefont{M.~J.} \bibnamefont{Puska}}, \bibnamefont{and}
  \bibinfo{author}{\bibfnamefont{R.~M.} \bibnamefont{Nieminen}},
  \bibinfo{journal}{Phys. Rev. B} \textbf{\bibinfo{volume}{54}},
  \bibinfo{pages}{15016} (\bibinfo{year}{1996}).

\bibitem[{\citenamefont{White and Bird}(1994)}]{Whi94}
\bibinfo{author}{\bibfnamefont{J.~A.} \bibnamefont{White}} \bibnamefont{and}
  \bibinfo{author}{\bibfnamefont{D.~M.} \bibnamefont{Bird}},
  \bibinfo{journal}{Phys. Rev. B} \textbf{\bibinfo{volume}{50}},
  \bibinfo{pages}{4954} (\bibinfo{year}{1994}).

\bibitem[{\citenamefont{Perdew et~al.}(1996)\citenamefont{Perdew, Burke, and
  Ernzerhof}}]{Per96}
\bibinfo{author}{\bibfnamefont{J.~P.} \bibnamefont{Perdew}},
  \bibinfo{author}{\bibfnamefont{K.}~\bibnamefont{Burke}}, \bibnamefont{and}
  \bibinfo{author}{\bibfnamefont{M.}~\bibnamefont{Ernzerhof}},
  \bibinfo{journal}{Phys. Rev. Lett.} \textbf{\bibinfo{volume}{77}},
  \bibinfo{pages}{3865} (\bibinfo{year}{1996}).

\bibitem[{\citenamefont{Kutzler and Painter}(1987)}]{Kut87}
\bibinfo{author}{\bibfnamefont{F.~W.} \bibnamefont{Kutzler}} \bibnamefont{and}
  \bibinfo{author}{\bibfnamefont{G.~S.} \bibnamefont{Painter}},
  \bibinfo{journal}{Phys. Rev. Lett.} \textbf{\bibinfo{volume}{59}},
  \bibinfo{pages}{1285} (\bibinfo{year}{1987}).

\bibitem[{\citenamefont{Kurth et~al.}(1999)\citenamefont{Kurth, Perdew, and
  Blaha}}]{Kur99}
\bibinfo{author}{\bibfnamefont{S.}~\bibnamefont{Kurth}},
  \bibinfo{author}{\bibfnamefont{J.~P.} \bibnamefont{Perdew}},
  \bibnamefont{and} \bibinfo{author}{\bibfnamefont{P.}~\bibnamefont{Blaha}},
  \bibinfo{journal}{Int. J. Quantum Chem.} \textbf{\bibinfo{volume}{75}},
  \bibinfo{pages}{889} (\bibinfo{year}{1999}).

\bibitem[{\citenamefont{Zhang and Yang}(1998)}]{Zha98}
\bibinfo{author}{\bibfnamefont{Y.}~\bibnamefont{Zhang}} \bibnamefont{and}
  \bibinfo{author}{\bibfnamefont{W.}~\bibnamefont{Yang}},
  \bibinfo{journal}{Phys. Rev. Lett.} \textbf{\bibinfo{volume}{80}},
  \bibinfo{pages}{890} (\bibinfo{year}{1998}).

\bibitem[{\citenamefont{Daykov et~al.}(2003)\citenamefont{Daykov, Arias, and
  Engeness}}]{Day03}
\bibinfo{author}{\bibfnamefont{I.~P.} \bibnamefont{Daykov}},
  \bibinfo{author}{\bibfnamefont{T.~A.} \bibnamefont{Arias}}, \bibnamefont{and}
  \bibinfo{author}{\bibfnamefont{T.~D.} \bibnamefont{Engeness}},
  \bibinfo{journal}{Phys. Rev. Lett.} \textbf{\bibinfo{volume}{90}},
  \bibinfo{pages}{216402} (\bibinfo{year}{2003}).

\bibitem[{\citenamefont{Perdew and Wang}(1992)}]{Per92}
\bibinfo{author}{\bibfnamefont{J.~P.} \bibnamefont{Perdew}} \bibnamefont{and}
  \bibinfo{author}{\bibfnamefont{Y.}~\bibnamefont{Wang}},
  \bibinfo{journal}{Phys. Rev. B} \textbf{\bibinfo{volume}{45}},
  \bibinfo{pages}{13244} (\bibinfo{year}{1992}).

\bibitem[{\citenamefont{Bahn and Jacobsen}(2002)}]{Bah02}
\bibinfo{author}{\bibfnamefont{S.~R.} \bibnamefont{Bahn}} \bibnamefont{and}
  \bibinfo{author}{\bibfnamefont{K.~W.} \bibnamefont{Jacobsen}},
  \bibinfo{journal}{Comp. in Sci. and Eng.} \textbf{\bibinfo{volume}{4}},
  \bibinfo{pages}{56} (\bibinfo{year}{2002}).

\bibitem[{\citenamefont{Hammer et~al.}(1999)\citenamefont{Hammer, Hansen, and
  N{\o}rskov}}]{Ham99}
\bibinfo{author}{\bibfnamefont{B.}~\bibnamefont{Hammer}},
  \bibinfo{author}{\bibfnamefont{L.~B.} \bibnamefont{Hansen}},
  \bibnamefont{and} \bibinfo{author}{\bibfnamefont{J.~K.}
  \bibnamefont{N{\o}rskov}}, \bibinfo{journal}{Phys. Rev. B}
  \textbf{\bibinfo{volume}{59}}, \bibinfo{pages}{7413} (\bibinfo{year}{1999}).

\bibitem[{Dac()}]{Dacapo}
\bibinfo{note}{The Dacapo code is freely available at
  http://www.fysik.dtu.dk/campos}.

\bibitem[{com()}]{complex}
\bibinfo{note}{There are 2897 complex plane-wave coefficients, but only half of
  the coefficients need to be stored, because the wave function can be chosen
  as real.}

\bibitem[{\citenamefont{Troullier et~al.}(1995)\citenamefont{Troullier,
  Chelikowsky, and Saad}}]{Tro95}
\bibinfo{author}{\bibfnamefont{N.}~\bibnamefont{Troullier}},
  \bibinfo{author}{\bibfnamefont{J.~R.} \bibnamefont{Chelikowsky}},
  \bibnamefont{and} \bibinfo{author}{\bibfnamefont{Y.}~\bibnamefont{Saad}},
  \bibinfo{journal}{Solid State Communications} \textbf{\bibinfo{volume}{93}},
  \bibinfo{pages}{225} (\bibinfo{year}{1995}).

\bibitem[{Gri()}]{GridPAW}
\bibinfo{note}{The code is written in the Python and C++ programming languages,
  and is freely available at http://www.fysik.dtu.dk/campos/gridpaw}.

\bibitem[{\citenamefont{Kerker}(1981)}]{Ker81}
\bibinfo{author}{\bibfnamefont{G.~P.} \bibnamefont{Kerker}},
  \bibinfo{journal}{Phys. Rev. B} \textbf{\bibinfo{volume}{23}},
  \bibinfo{pages}{3082} (\bibinfo{year}{1981}).

\bibitem[{\citenamefont{Obara and Saika}(1986)}]{Oba86}
\bibinfo{author}{\bibfnamefont{S.}~\bibnamefont{Obara}} \bibnamefont{and}
  \bibinfo{author}{\bibfnamefont{A.}~\bibnamefont{Saika}}, \bibinfo{journal}{J.
  Chem. Phys} \textbf{\bibinfo{volume}{84}}, \bibinfo{pages}{3963}
  (\bibinfo{year}{1986}).

\end{thebibliography}

\end{document}